\pdfoutput=1
\documentclass[11pt,a4paper]{article}
\usepackage[margin=1in]{geometry}
\usepackage{mathtools}
\usepackage{amsmath}
\usepackage{amsfonts}
\usepackage{amsthm}
\usepackage{nicefrac}
\usepackage{algorithmic}
\usepackage[linesnumbered,lined,boxed,ruled,vlined,noend]{algorithm2e}
\SetKwInput{KwInput}{Input}
\SetKwInput{KwOutput}{Output}
\SetKwInOut{Parameters}{Parameters}
\SetKwComment{Comment}{$\triangleright$\ }{}
\SetAlFnt{\small}
\SetAlCapFnt{\small}
\SetAlCapNameFnt{\small}
\SetAlCapHSkip{0pt}
\IncMargin{-\parindent}

\usepackage{microtype}

\RequirePackage[dvipsnames, table, hideerrors]{xcolor} %
\definecolor{LightBlue}{rgb}{0.5, 0.6, 0.9}
\definecolor{DarkGreen}{rgb}{0.1,0.5,0.1}
\definecolor{one}{RGB}{51,34,136}
\definecolor{two}{RGB}{17,119,51}
\definecolor{three}{RGB}{136,34,85}
\usepackage{bbm}
\usepackage{color}
\usepackage{enumitem}
\usepackage[dvipsnames]{xcolor}
\usepackage{thmtools}
\usepackage{caption}
\usepackage{booktabs}
\usepackage{multirow}

\usepackage{hyperref}
\usepackage[capitalize,nameinlink]{cleveref}
\hypersetup{colorlinks,linkcolor={blue},citecolor={blue},urlcolor={red}}

\newtheorem{theorem}{Theorem}[section]
\newtheorem{lemma}[theorem]{Lemma}

\newtheorem{fact}{Fact}[section]
\theoremstyle{definition}
\newtheorem{definition}[theorem]{Definition}
\newtheorem{invariant}[theorem]{Invariant}

\newtheorem{observation}[theorem]{Observation}

\newcommand{\whp}{\emph{whp}\xspace}
\newcommand{\eps}{\varepsilon}
\newcommand{\defn}[1]{\textbf{\emph{#1}}}
\newcommand{\etal}{et al.\xspace{}}

\newcommand\numberthis{\addtocounter{equation}{1}\tag{\theequation}}

\newcommand{\E}{\ensuremath{\mathop{{}\mathbb{E}}}}

\usepackage{bbm}

\newcommand{\apx}[1]{\ensuremath{\widetilde{#1}}}

\newcommand{\density}{\ensuremath{\rho}}
\newcommand{\abs}[1]{\ensuremath{\left| #1 \right|}}

\newcommand{\Sout}{\ensuremath{S_{\mathrm{out}}}}

\newcommand{\core}{k}

\newcommand{\kest}{\hat{\core}}

\newcommand{\lf}{\psi}
\newcommand{\upexp}{(1+\lf)}
\newcommand{\upexpold}{(2+\lambda)(1+\lf)}

\newcommand{\hnup}{\widehat{\nup}}

\newcommand{\lcur}{r}

\newcommand{\release}{\textbf{releases}\xspace}

\newcommand{\rangeout}{\mathcal{Y}}
\newcommand{\adj}{\mathbf{a}}
\newcommand{\coren}{\eta}

\newcommand{\numlevels}{4\log^2 n}
\newcommand{\numgrouplevels}{2\log n}

\newcommand{\nup}{\mathcal{U}}

\newcommand{\alg}{\mathcal{A}}

\newcommand{\floor}[1]{\lfloor #1 \rfloor}
\newcommand{\ceil}[1]{\lceil #1 \rceil}

\newcommand{\poly}{\text{poly}\xspace}
\newcommand{\kc}{$k$-core\xspace}

\newcommand{\Lap}{\ensuremath{\mathsf{Lap}}}
 
\crefname{theorem}{Theorem}{Theorems}
\crefname{lemma}{Lemma}{lemmas}
\Crefname{lemma}{Lemma}{Lemmas}
\Crefname{claim}{Claim}{Claims}
\crefname{invariant}{Invariant}{Invariants}
\Crefname{observation}{Observation}{Observations}
\Crefname{algorithm}{Algorithm}{Algorithms}
\Crefname{myalgctr}{Algorithm}{Algorithms}
\Crefname{challenge}{Challenge}{Challenges}
\crefalias{AlgoLine}{line}

\crefname{algocf}{alg.}{algs.}
\Crefname{algocf}{Algorithm}{Algorithms}

\title{Near-Optimal Differentially Private Graph Algorithms via the Multidimensional AboveThreshold Mechanism}
\date{}
\author{Laxman Dhulipala$^1$, Monika Henzinger$^2$, George Z. Li$^3$,\\ Quanquan C. Liu$^4$, A.\ R.\ Sricharan$^5$, and Leqi Zhu$^6$}
\date{$^1$University of Maryland, $^2$Institute of Science and Technology, Austria,\\ $^3$Carnegie Mellon University, $^4$Yale University, $^5$UniVie Doctoral School Computer Science DoCS, University of Vienna, Austria, $^6$University of Manitoba}

\begin{document}
\maketitle
\date{}

\begin{abstract}
Many differentially private and classical non-private graph algorithms rely crucially on determining whether some property of each vertex meets a threshold. For example, for the $k$-core decomposition problem, the classic peeling algorithm iteratively removes a vertex if its induced degree falls below a threshold. The sparse vector technique (SVT) is generally used to transform non-private threshold queries into private ones with only a small additive loss in accuracy. However, a naive application of SVT in the graph setting leads to an amplification of the error by a factor of $n$ due to composition, as SVT is applied to \emph{every vertex}. In this paper, we resolve this problem by formulating a novel \emph{generalized} sparse vector technique which we call the \emph{Multidimensional AboveThreshold (MAT) Mechanism} which generalizes SVT (applied to vectors with one dimension) to vectors with \emph{multiple} dimensions. When applied to vectors with $n$ dimensions, we solve a number of important graph problems with better bounds than previous work. 

Specifically, we apply our MAT mechanism to obtain a set of improved bounds for a variety of problems including $k$-core decomposition, densest subgraph, low out-degree ordering, and vertex coloring. We give a \emph{tight} local edge differentially private (LEDP) algorithm for $k$-core decomposition that results in an approximation with $O(\eps^{-1}\log n)$ additive error and \emph{no} multiplicative error in $O(n)$ rounds. We also give a new $(2+\eta)$-factor multiplicative, $O(\eps^{-1}\log n)$ additive error algorithm in $O(\log^2 n)$ rounds for any constant $\eta > 0$. Both of these results are asymptotically tight against our new lower bound of $\Omega(\log n)$ for any constant-factor approximation algorithm for $k$-core decomposition. Our new algorithms for $k$-core decomposition also directly lead to new algorithms for the related problems of densest subgraph and low out-degree ordering. Finally, we give novel LEDP differentially private defective coloring algorithms that use number of colors given in terms of the \emph{arboricity} of the graph. 
\end{abstract}

\section{Introduction}

Designing and analyzing algorithms that satisfy \emph{differential privacy}~\cite{DMNS06}, the de facto standard for protecting user data, is a tricky task, with few general techniques. Among them, the \emph{sparse vector technique (SVT)} is a well-known approach to make certain \emph{threshold-based} computations privacy-preserving. This typically involves showing that the computation can be reduced to a sequence of black-box invocations of private \emph{AboveThreshold} mechanisms, followed by some post-processing.
Each such mechanism allows one to iteratively (and adaptively) query whether some real-valued property of the data is above some threshold, until the first ``yes'' response, at which point no further queries are possible.

While SVT was originally presented as a technique for \emph{sequential} computations, recent work on concurrent composition~\cite{VW21concurrent} show that it is possible to interleave queries to different (AboveThreshold) mechanisms with (worst-case) privacy loss proportional to the number of concurrently queried mechanisms. However, this can be prohibitively large for applications (e.g., in parallel and distributed settings) where there is a large amount of concurrency and, in this case, one must exploit structural properties of the specific instance to obtain better guarantees.

To analyze these situations, we introduce a new mechanism, called \emph{Multidimensional AboveThreshold (MAT)}, which allows one to query, \emph{in parallel}, whether a number of different real-valued properties of the data are above some specified thresholds. We identify a condition on the sensitivity of the submitted queries that allows one to use MAT with far less privacy loss versus concurrent composition. By analyzing various non-private algorithms through the lens of our mechanism, we obtain private algorithms with substantially improved accuracy guarantees for a variety of important, recently studied graph problems.

All of our algorithms apply in the stringent, \emph{local} model of (edge) differential privacy \emph{(LEDP)}, where each user (here, a vertex in the graph) interactively discloses information about its data (here, its incident edges) in multiple \emph{rounds} of communication with an \emph{untrusted}\footnote{The curator may accidentally leak their messages, e.g., due to security breaches. However, it is not malicious. This is referred to as ``honest but curious'' in the literature.} curator (or server), which produces the output. The \emph{transcript} of messages sent throughout the interaction must be differentially private.

We now describe our contributions and the main related work in more detail (see~\autoref{tab:results} for a summary of our results and see the appendix for more related work). In the following, all graphs are simple (no self-loops), undirected, and contain $n$ vertices.
Furthermore, all mentioned additive error upper bounds hold \emph{with high probability}, i.e., $1-1/n^c$, for any constant $c>0$.

\paragraph{$k$-Core Decomposition.}
At a high level, core numbers are a measure of the relative importance of vertices in a graph. Formally, given a graph $G$, the {\em core number} (or, \emph{coreness}) of a vertex $v$ in $G$ is the largest integer $k_G(v)$ (or simply $k(v)$, if $G$ is clear from the context) for which there exists a subgraph $H$ of $G$ \emph{containing $v$} such that every vertex in $H$ has induced degree at least $k_G(v)$. In the \emph{$k$-core decomposition problem}, the goal is to output the core numbers of all vertices in a given graph\footnote{We note that the $k$-core decomposition problem in the literature is sometimes used to refer to the problem of computing a hierarchy of \emph{$k$-cores}, which are subgraphs of $G$ induced by vertices with core number at least $k$~\cite{sariyuce2016fast}; note that it is not possible to emit such a hierarchy privately, so we (and prior work) focus on privately emitting core numbers for every vertex.}.

This problem has recently received much attention
from the database and systems communities due its applications in community detection~\cite{Alvarez2005,Esfandiari2018,Ghaffari2019}, clustering and data mining~\cite{costa2011analyzing, shin2016corescope},
and other machine learning and graph analytic applications~\cite{Kabir2017, dhulipala2017julienne, dhulipala2018theoretically}; for a more comprehensive survey, see~\cite{malliaros2020core}. Due to the rapid increase of applications using sensitive user data,
designing privacy-preserving algorithms for this problem is becoming increasingly important.

An algorithm is
a \emph{$(\phi, \zeta)$-approximation} (or, a \emph{$\phi$-approximate algorithm with additive error $\zeta$}) for the $k$-core decomposition problem if, for every vertex $v$ in the given graph, the \emph{estimate} of the core number of $v$, as returned by the algorithm, lies between $k(v) - \zeta$ and $\phi\cdot k(v) + \zeta$.
Dhulipala \etal~\cite{DLRSSY22} gave the first differentially private $(2+\eta, O(\eps^{-1}\log^3 n))$-approximation algorithm  for $k$-core decomposition in the $\eps$-LEDP model, which runs in $O(\log n)$ rounds. They left it as an open question whether their multiplicative and additive approximation
factors can be improved.

\medskip

\emph{Our contributions.} We answer their question in the affirmative. In particular, we  give an \emph{exact} (i.e., 1-approximate) algorithm and a $(2+\eta)$-approximate algorithm, both with $O(\eps^{-1}\log n)$ additive error, which run in $O(n)$ and $O(\log^2 n)$ rounds, respectively.

\begin{theorem}\label{thm:kcore-n-round}
For $\eps > 0$, there is an exact, $O(n)$-round $\eps$-LEDP algorithm for the $k$-core decomposition problem, which has $O(\eps^{-1}\log n)$ additive error, with high probability.
\end{theorem}

\begin{theorem}\label{thm:small-rounds-kcore}
    For $\eps > 0$ and \emph{constant} $\eta > 0$, there is a $(2+\eta)$-approximate, $O(\log^2 n)$-round $\eps$-LEDP algorithm for the $k$-core decomposition problem, which has  $O(\eps^{-1}\log n)$ additive error, with high probability.
\end{theorem}

Furthermore, we show that the additive errors of our algorithms are tight, by proving a matching lower bound of $\Omega(\eps^{-1}\log n)$ on the additive error of any constant-approximation algorithm in the \emph{centralized} model (EDP), where the algorithm has full access to user data (here, the input graph). Since the centralized model is less restricted than the local model, the same bound holds in the LEDP model, regardless of round complexity. We also show that interactivity is needed if one wants an exact algorithm with polylogarithmic additive error, by proving a lower bound of $\Omega(\sqrt{n})$ on the additive error of any exact, 1-round algorithm, when $\eps$ is a constant.

\begin{theorem}
\label{thm:centrallb}
    For $\eps > 0$ and $\phi \geq 1$, any $\phi$-approximate $\eps$-EDP algorithm for the $k$-core decomposition problem in the centralized model has $\Omega((\eps \phi)^{-1} \log n)$ additive error, with constant probability.
\end{theorem}

\begin{theorem}
\label{thm:oneroundlb}
For \emph{constant} $\eps > 0$, any 1-round, exact $\eps$-LEDP algorithm for the $k$-core decomposition problem has $\Omega(\sqrt{n})$ additive error, with constant probability.
\end{theorem}

\paragraph{Densest Subgraph.}
Another measure of importance, defined on \emph{subsets} of vertices in a graph, is density. Formally, the \emph{density} of a set of vertices $S$ in a graph $G$, denoted $\rho_G(S)$ (or simply $\rho(S)$, if $G$ is clear from the context), is the ratio of the number of edges in the subgraph induced by $S$ and the number of vertices in $S$, i.e., $|E(S)|/|S|$. In the \emph{densest subgraph} problem, the goal is to output a set of vertices with the maximum density in a given graph.

This problem has been studied in the (centralized) private \cite{AHS21,NV21},
LEDP \cite{DKLV23,DLRSSY22},
dynamic \cite{DBLP:conf/stoc/BhattacharyaHNT15,sawlani2020near,chekuri2024adaptive}, streaming \cite{BGM14,BHNT15,esfandiari2016brief,mcgregor2015densest},
and distributed~\cite{BKV12,SuVu20} settings; for a recent survey on densest subgraph and its variants, see~\cite{lanciano2023survey}.

An algorithm is a \emph{$(\phi,\zeta)$-approximation} to the densest subgraph problem if the set of vertices $S$ returned by the algorithm has density $\rho(S) \geq \rho^*/\phi - \zeta$, where $\rho^*$ is the maximum density. In the $\eps$-LEDP setting, Dhulipala \etal~\cite{DLRSSY22} gave a $(4+\eta,O(\eps^{-1}\log^3 n))$-approximation for the densest subgraph problem. More recently, and concurrently with this work, Dinitz \etal~\cite{DKLV23} gave a $(2+\eta, O((\eps\eta)^{-1}\log^2 n))$-approximation algorithm, which runs in $O(\log n)$ rounds, in the $\eps$-LEDP setting, and as well as an $(1,O(\eps^{-1}\sqrt{\log \delta^{-1}}\log n))$-approximation, which runs in $O(n^2\log n)$ rounds in the $(\eps,\delta)$-LEDP setting. On the other hand, there is a lower bound of $\Omega(\phi^{-1}\sqrt{\eps^{-1}\log n})$ on the additive error of any $\phi$-approximate algorithm for densest subgraph that satisfies $(\eps,\delta)$-edge differential privacy in the centralized model~\cite{AHS21,NV21} (when $\delta \leq n^{-O(1)}$).

\medskip

\emph{Our contributions.} We give a condition on when a (private) $(\phi,\zeta)$-approximation algorithm for the $k$-core decomposition problem can be transformed into a (private) $(2\phi,\zeta)$-approximation algorithm for the densest subgraph problem. We show that our algorithms for $k$-core decomposition satisfy the condition, hence obtaining the following improved bounds for densest subgraph.

\begin{theorem}\label{thm:ds}
For $\eps > 0$, there is a $2$-approximate, $O(n)$-round $\eps$-LEDP algorithm for the densest subgraph problem, which has  $O(\eps^{-1}\log n)$ additive error, with high probability.
\end{theorem}

\begin{theorem}\label{thm:small-rounds-ds}
For $\eps > 0$ and \emph{constant} $\eta > 0$, there is a $(4+\eta)$-approximate, $O(\log^2 n)$-round $\eps$-LEDP algorithm for the densest subgraph problem, which has $O(\eps^{-1}\log n)$ additive error, with high probability.
\end{theorem}

We also give a simple 1-round algorithm using the randomized response technique together with brute-force, exponential time search. We believe this algorithm serves as an interesting example of what a lower bound for 1-round algorithms must be able to reason about.

\begin{theorem}\label{thm:one-round-ds}
For $\eps > 0$, there is an exact, 1-round $\eps$-LEDP algorithm for the densest subgraph problem, which has $O((1 + \eps^{-1})\sqrt{n})$ additive error, with high probability.
\end{theorem}

\paragraph{Low Out-Degree Ordering.}

A \emph{low out-degree orientation} of a (undirected) graph is an orientation of the edges of the graph such that the out-degree of any vertex is minimized. A  \emph{low out-degree ordering} is a permutation of the vertices, which implicitly encodes a low out-degree orientation, i.e., each edge is oriented from its earlier endpoint to its later endpoint (with respect to the permutation). In the \emph{low out-degree ordering problem}, the goal is to output a low out-degree ordering of a given graph.

Differentially private low out-degree ordering was introduced by~\cite{DLRSSY22} as a way to release an implicit solution to the low out-degree orientation problem. The latter has attracted a lot of attention in the non-private setting (see, e.g.,~\cite{kowalik2006approximation,SuVu20,berglin2020simple,chekuri2024adaptive} and references
therein) and has been used to improve the running time of algorithms for  matching~\cite{neiman2015simple,he2014orienting},
coloring~\cite{HNW20,ghaffari2024dynamic}, and subgraph counting~\cite{BGLSS22,BeraPS20,LSYDS22}.

The \emph{degeneracy} of a graph is the maximum core number of any vertex in the graph. It is known that degeneracy is equal to the minimum out-degree that is achievable through any ordering. Hence, we state the out-degree obtained through our algorithms in terms of $\alpha$, the degeneracy
of the input graph.
An algorithm is a \emph{$(\phi,\zeta)$-approximation} for the low out-degree ordering problem if the output of the algorithm encodes an orientation where each vertex has out-degree at most $\phi\cdot\alpha + \zeta$.
In the $\eps$-LEDP model, Dhulipala \etal~\cite{DLRSSY22} gave a $(2+\eta, O(\eps^{-1}\log^3 n))$-approximation to the low out-degree ordering problem, which runs in $O(\log n)$ rounds.

\emph{Our contributions.} We improve the multiplicative factor and the additive error by giving an exact algorithm and a $(2+\eta)$-approximate algorithm, both with additive error $O(\eps^{-1}\log n)$, which run in $O(n)$ and $O(\log n)$ rounds, respectively.

\begin{theorem}\label{thm:low-out-degree-n-round}
    For $\eps > 0$, there is an exact, $O(n)$-round $\eps$-LEDP algorithm algorithm for the
    low out-degree ordering problem, which has $O(\eps^{-1}\log n)$ additive error, with high probability.
\end{theorem}

\begin{theorem}\label{thm:low-out-degree-low-round}
    For $\eps > 0$ and \emph{constant} $\eta > 0$, there is a  $(2+\eta)$-approximate, $O(\log^2 n)$-round $\eps$-LEDP algorithm for the
    low out-degree ordering problem, which has $O(\eps^{-1}\log n)$ additive error, with high probability.
\end{theorem}

\paragraph{Defective Coloring.}
     A (vertex) \emph{coloring} of a graph $G$ is an assignment of labels (or, \emph{colors}) from the set $\{1,\ldots,n\}$ to the vertices of $G$. A coloring is \emph{$(c,d)$-defective} if it uses at most $c$ different colors, and each vertex has at most $d$ neighbors with the same color; $d$ is the \emph{defectiveness} of the coloring. In the \emph{defective coloring} problem, the goal is to output a $(c,d)$-defective coloring of a given graph with small $c$ and $d$.

   This problem (especially the case $d = 0$) has been widely studied in a variety of settings including the (centralized)  private~\cite{christiansen2024private},
dynamic~\cite{BGKLS22,HNW20,HP22,SW20}, distributed~\cite{ghaffari2022deterministic,GL17,Maus23},
and streaming~\cite{ACGS23,BBMU21,BCG20} settings. In the $\eps$-edge differential privacy setting, \cite{christiansen2024private} recently gave the first $(c,d)$-defective coloring algorithm with $c = O(\Delta/\log n + \eps^{-1})$ and $d = O(\log n)$, where $\Delta$ is the maximum degree of the input graph. They also showed that $d = \Omega(\log n/(\log c + \log \Delta ))$ defectiveness
is necessary if at most $c$ colors are used. Although not explicitly
stated, we believe their coloring algorithm can be also adapted
to the $\eps$-LEDP setting with the same bounds.

\emph{Our contributions.}
Using the approximate low out-degree ordering results and MAT, we give new algorithms for defective coloring under LEDP. In particular, the number of colors used by our algorithms scales with the arboricity $\alpha$ instead of the maximum degree. In small-arboricity graphs, the number of colors used by our algorithm can be significantly smaller than
that  of~\cite{christiansen2024private}.

\begin{theorem}
\label{thm:defective-n-round}
     For $\eps > 0$, there is an $O(n$)-round $\eps$-LEDP algorithm for the defective coloring problem that uses $O(1 + \alpha\eps/\log{n})$ colors, where $\alpha$ is the arborocity of the input graph, and has $O(\eps^{-1}\log{n}))$ defectiveness, with high probability.
\end{theorem}

\begin{theorem}
\label{thm:defective-small-round}
     For $\eps > 0$, there is an $O(\log^2 n)$-round $\eps$-LEDP algorithm for the defective coloring problem that uses $O(\alpha\eps\log{n} + \log^2 n)$ colors, where $\alpha$ is the arborocity of the input graph, and has $O(\eps^{-1}\log{n}))$ defectiveness, with high probability.
\end{theorem}

\paragraph{Efficient centralized algorithms.} All of the bounds in the above theorems are stated for $\eps$-LEDP algorithms but also are algorithms in the centralized model with the same approximation and additive error guarantees. All of our algorithms based on MAT can be implemented in the centralized model in near-linear time if we lose an additional multiplicative factor of $1+\eta$ for any given constant $\eta>0$. This nearly matches the best running times for
sequential non-private $k$-core decomposition algorithms, while maintaining a near-optimal utility for private algorithms.

\paragraph{Concurrent, Independent Work.} The recent concurrent, independent work of Dinitz et al.~\cite{DKLV23} provides a large set of
novel results on $(\eps, \delta)$-DP, $\eps$-LEDP, and $(\eps, \delta)$-LEDP densest subgraph, improving on the approximation guarantees of
previous results. Their work is based on a recent result of Chekuri, Quanrud, and Torres~\cite{CQT22} and the parallel densest subgraph
algorithm of Bahmani, Kumar, and Vassilvitskii~\cite{BKV12}. They offer improvements in the bounds in the central DP setting via the
private selection mechanism of Liu and Talwar~\cite{liu2019private}. %
Note that their techniques are different from those presented in this paper; furthermore, any $c$-approximate
algorithm for densest subgraphs translates to a $2c$-approximate algorithm for finding the degeneracy of the input graph (which is the value
of the maximum $k$-core). Thus, no densest subgraph algorithm can obtain a better than $2$-approximation of the maximum $k$-core value.

\paragraph{Subsequent Work.} Using the MAT framework of this work, Dinitz et al.~\cite{dinitz2024private} gave new algorithms for node-differentially private bipartite matching, improving upon the results of~\cite{hsu2014private}. They also gave the first algorithms for maximum matchings in general graphs under local edge-differential privacy, also crucially using the MAT framework. Complementing their algorithms, they show a matching lower bound for edge-private maximum matchings. This follow up work highlights the wide applicability of the MAT framework, giving nearly tight algorithms for various combinatorial problems under differential privacy.

\bigskip

\begin{table}[!th]
\footnotesize
\centering
\begin{tabular}{cccccc}
\toprule
\textbf{Problem} & \textbf{Apx. Factor} & \textbf{Additive Error}  & \textbf{Rounds} &  \textbf{Citation}  \\
\midrule
\multirow{5}{*}{\shortstack{\textbf{Core} \\ \textbf{Decomposition}}} & $1$ & $O(\eps^{-1}\log n)$ & $O(n)$ &  \cref{thm:kcore-n-round} \\
& $1$ & $\Omega(\sqrt{n})$ (constant $\eps$) & 1 &  \cref{thm:oneroundlb}  \\
& $2+\eta$ & $O(\eps^{-1}\log n)$ & $O(\log^2 n)$ &  \cref{thm:small-rounds-kcore} \\
& $2+\eta$ & $O(\eps^{-1}\log^3 n)$ & $O(\log n)$ &  \cite{DLRSSY22}  \\
& $\phi$ & $\Omega((\eps\phi)^{-1}\log n)$ & any &  \cref{thm:centrallb} \\ \midrule
\multirow{6}{*}{\shortstack{\textbf{Densest} \\  \textbf{Subgraph}}}
& $1$ & $O(\eps^{-1}\sqrt{\log (\delta^{-1})}\log n)$ & $O(n^2\log n)$ & \cite{DKLV23}$^*$  \\
& $1$ & $O((1 + \eps^{-1})\sqrt{n})$ & 1 & \cref{thm:one-round-ds} \\
& $\phi$ & $\Omega(\phi^{-1}\sqrt{\eps^{-1} \log n})$ & any & \cite{AHS21,NV21}\\
& $2$ & $O(\eps^{-1}\log n)$ & $O(n)$ &  \cref{thm:ds} \\
& $2+\eta$ & $O(\eps^{-1}\eta^{-1}\log^2 n)$ & $O(\eta^{-1}\log n)$ & \cite{DKLV23}  \\
& $4+\eta$ & $O(\eps^{-1}\log^3 n)$ & $O(\log n)$ & \cite{DLRSSY22}  \\
& $4+\eta$ & $O(\eps^{-1}\log n)$ & $O(\log^2 n)$ & \cref{thm:small-rounds-ds} \\
\midrule
\multirow{3}{*}{\shortstack{\textbf{Low} \\ \textbf{Out-degree} \\ \textbf{Ordering}}} & $1$ & $O(\eps^{-1}\log n)$ & $O(n)$ &  \cref{thm:low-out-degree-n-round} \\
& $2+\eta$ & $O(\eps^{-1}\log n)$ & $O(\log^2 n)$ &  \cref{thm:low-out-degree-low-round} \\
& $2+\eta$ & $O(\eps^{-1}\log^3 n)$ & $O(\log n)$ &  \cite{DLRSSY22}\\
\midrule
& \textbf{\# Colors} & \textbf{Defectiveness}  &  &   \\
\midrule
\multirow{4}{*}{\shortstack{\textbf{Defective} \\ \textbf{Coloring}}} & $O(1+\eps\alpha/\log{n})$ & $O(\eps^{-1}\log n)$ & $O(n)$ &  \cref{thm:defective-n-round} \\
& $O(\eps\alpha\log{n} + \log^2 n)$ & $O(\eps^{-1}\log n)$ & $O(\log^2 n)$ &  \Cref{thm:defective-small-round}\\
& $O(\eps^{-1}+\Delta/\log{n})$ & $O(\log n)$ & 2 &  \cite{christiansen2024private}
\\
& $c$ & $\Omega(\frac{\log(n)}{\log{c}+\log{\Delta}})$ & any &  \cite{christiansen2024private} \\
\bottomrule
\end{tabular}
\caption{Summary of error bounds in the local model. Each upper bound is for a mechanism satisfying $(\eps,\delta)$-edge differential privacy and holds with high probability. An asterisk in the citation indicates that an algorithm is $(\eps,\delta)$-LEDP instead of $\eps$-LEDP. Lower bounds are for $\eps$-differential privacy and hold with constant probability.}
\label{tab:results}
\end{table}

\section{Technical Overview}

\paragraph{MAT.} Our new \emph{Multidimensional
AboveThreshold} (MAT) mechanism generalizes the AboveThreshold mechanism~\cite{dwork2009complexity} to support a $d$-dimensional vector of queries. Conceptually, one may think of MAT as running $d$ copies of AboveThreshold in parallel, one per dimension. More precisely, suppose that the input dataset for (single-dimensional) AboveThreshold was from some domain $\mathcal X$, then the input for our mechanism is $X = (x_1,\dots,x_d) \in \mathcal{X}^d$, a $d$-tuple of elements from $\mathcal{X}$. Our mechanism takes as input an adaptively chosen sequence $\vec{f}_1,\vec{f}_2,\dots$ of \emph{query vectors} and a \emph{threshold vector} $\vec{T} = (T_1,\dots,T_d)$. Each query vector specifies $d$ \emph{queries}, one for each dimension of the dataset, i.e., $\vec{f}_i = (f_{i,1},\dots,f_{i,d})$ and  query $f_{i,j} : \mathcal{X} \to \mathbb{R}$ is on element $x_j$.
The goal is to (privately) return, after each query vector $\vec{f}_i$, a response vector $\vec{a}_i \in \{\bot,\top\}^d$ indicating whether each query $f_{i,j}$ in the vector has \emph{crossed corresponding threshold} $T_j$, i.e., $\vec{a}_i = (a_{i,1},\dots,a_{i,d})$ and $a_{i,j}$ indicates whether $f_{i,j}(x_j) \geq T_j$ ($\bot$ for ``no'', $\top$ for ``yes''). After some query $f_{i,j}$ has crossed its corresponding threshold, then all future queries $f_{i',j}$, for $i' > i$, on element $x_j$ are ignored (e.g., the response is always $\bot$). It is assumed that new query vectors are submitted until each element has had a query that has crossed its corresponding threshold. In a nutshell, MAT allows one to infer, for each element $x_j$, an estimate $\tilde{r}(j)$ of the index $r(j)$ of the \emph{first} query $f_{r(j),j}$ on $x_j$ that crosses the threshold $T_j$.

\paragraph{Sensitivity condition.}

We define a fairly technical condition on the \emph{sensitivity} of the queries, generalizing the $\ell_1$ sensitivity in the $1$-dimensional case, such that the error of MAT can be bounded in terms of the sensitivity.
For intuition, we first describe a simple case of our sensitivity condition.

Suppose that two datasets $X = (x_1,\dots,x_d)$ and $X' = (x_1',\dots,x_d')$ are neighboring only if they differ on at most $k \ll d$ coordinates, i.e., there exists a set $S \subseteq [d]$ of at most $k$ indices such that $x_i = x_i'$ for all $i \notin S$ and $x_i \sim x_i'$ for all $i \in S$.  Furthermore, suppose that the absolute difference of each query (on the appropriate element in $X$ and $X'$) is bounded by 1, i.e., for any query vector $\vec{f} = (f_1,\dots,f_d)$, each query $f_j$ satisfies $|f_j(x_j) - f_j(x_j')| \leq 1$. Since MAT runs $d$ copies of AboveThreshold in parallel and the queries are adaptive, applying concurrent composition~\cite{VW21concurrent} in this case would predict that the privacy loss is proportional to $d$. However, by coupling the outputs of the queries on the indices \emph{not} in $S$ (where $X$ and $X'$ are identical), it seems fairly intuitive that the actual privacy loss \emph{should} be proportional to $k$.
Hence, as the error of AboveThreshold is inversely proportional to the privacy loss parameter, to obtain $\eps$-differential privacy, we \emph{should} have at most $k \ll d$ times the original error of AboveThreshold in this setting, i.e., $O(k\eps^{-1}\log n)$ with high probability.

More generally, for an arbitrary neighboring relation  and arbitrary queries, roughly speaking, we define the sensitivity $D$ as the worst-case (over pairs of neighboring datasets $X$ and $X'$) sum of the maximum absolute difference in the outputs of the queries  (on $X$ and $X'$); see~\Cref{sec:mat} for a formal definition. In particular, for the simple case described above, $D = k$.
Using this definition, we formalize the above intuition and prove that MAT guarantees an error of $O(D\eps^{-1}\log n)$ in our applications, with high probability.

\paragraph{Applications of MAT.} We apply MAT to get new private algorithms for the $k$-core decomposition, densest subgraph,
low out-degree ordering, and defective coloring problems. In each case, given a graph $G = (V,E)$ with $n$ vertices, we show that some non-private algorithm can be simulated via MAT, for suitable definitions of ``dataset'' and ``queries''.

As a simple example, for the core decomposition problem, a dataset $X = (x_u : u \in V)$ is an $n$-tuple containing the set of neighbors $x_u$ of each vertex $u$ in $G$. (Hence, the domain $\mathcal{X}$ is the set of all subsets of $V$.) A query $f_S : \mathcal{X} \to \mathbb{R}$ counts the number of neighbors of the vertex that are \emph{not} among some fixed set of vertices $S \subseteq V$, shifted by some fixed amount $i$, and ``mirrored'' by $n$, i.e., $f_S(x) = n-(|x \setminus S| - i)$. (Notice that $f_S(x_u) \geq n$ means that vertex $u$ has at most $i$ neighbors that are not among $S$.) In this case, the absolute difference of each query on two neighboring datasets (i.e., graphs that differ by at most an edge) is at most 1 and there are at most two elements that differ for two neighboring datasets (namely, at the endpoints of the differing edge). Hence, the simple case mentioned above applies and MAT guarantees an error of $O(\eps^{-1}\log n)$, with high probability.

For core decomposition, we apply MAT to analyze a variant of the
classic peeling algorithm of Matula and Beck~\cite{MB83}, which iteratively peels (removes) the lowest degree vertex in the graph.
In our variant of the classical algorithm, we view the algorithm as running $n$ iterations of a \emph{peeling process}. In the $i$th iteration, the algorithm repeatedly peels all vertices with (induced) degree less than $i$ until none are left. It is known that, at the end of the $i$th iteration, a vertex is alive (has not been peeled) if and only if it is in the $i$-core. It can be shown that the peeling process can be simulated by MAT via the queries mentioned above. Hence, with some post-processing, we can obtain core
numbers with no multiplicative factor and the optimal (up to a constant factor
against our lower bound) additive error of $O(\eps^{-1}\log n)$, with high probability. Similarly, we  apply MAT to analyze the more round-efficient $(2+\eta)$-approximate
core decomposition  algorithm of~\cite{DLRSSY22}, obtaining optimal (up to a constant factor) error, with high probability.

Using our new core decomposition algorithms, we achieve improved additive error bounds for both the densest subgraph and low out-degree ordering
problems. In particular, we show that the $2$-approximate densest subgraph with additive error $O(\eps^{-1}\log n)$ is contained in the
subgraph consisting of all vertices with the maximum estimated core numbers (as we run the peeling algorithm, it can be shown that these vertices necessarily induce a subgraph with sufficiently high minimum degree).
Furthermore, all of our core decomposition algorithms directly return a low out-degree ordering with the same multiplicative and additive errors as the
original core decomposition algorithms by taking the order in which the nodes are peeled.

We also apply MAT to defective coloring by using MAT to greedily pick colors for each vertex
from the available colors that have not yet been occupied by too many neighbors. In particular, each dimension represents a
vertex and color pair consisting of all vertices in the graph and all available colors.
For dimension $j$ representing pair $(v, c)$, the $j$-th dimension contains queries asking whether
the number of neighbors of $v$ colored with $c$ exceeded our defectiveness threshold $O(\eps^{-1}\log n)$. If it has, we will
remove $c$ from $v$'s available set of colors. Our low round coloring algorithm follows a similar structure, and is inspired by the non-private algorithm of~\cite{HNW20}
and uses our low out-degree ordering algorithm to implement a separate set of available colors
for each of $O(\log^2 n)$ groups of vertices of similar core numbers.

\paragraph{Local lower bound.} Our lower bound for 1-round, exact core decomposition in the local model is obtained by reduction to a problem in the \emph{centralized} model, for which there is a strong lower bound. Specifically, it is known that, to privately answer $\Theta(n)$ \emph{random} inner product queries on a secret $\{0,1\}^n$ vector, most responses need to have $\Omega(\sqrt{n})$ additive error~\cite{de2012lower}. Recently, \cite{ELRS22} gave an elegant lower bound on the additive error of 1-round triangle counting algorithms in the local model using similar techniques.

In our case, we define a class of ``query graphs'', one per inner product query, in which the core number of a \emph{fixed} vertex $x$ is (roughly) the answer to the query.
In addition to $x$, there are some \emph{secret} vertices (and other vertices). For each secret vertex $v$, the existence of the edge $\{v,x\}$ is private information (which depends on the secret vector).
Crucially, all neighborhoods of $x$ and the secret vertices in {\em any possible query graph}  %
appear in two specific query graphs, namely, ones corresponding to the all-ones and all-zeros query vectors, respectively. The neighborhoods of the remaining vertices do not depend on the secret vector.

Our approach to solving the inner product problem is now as follows. The centralized algorithm first simulates the 1-round local core decomposition algorithm on the two fixed graphs to determine the messages that $x$ and the secret vertices would send in {\em any} query graph and saves these messages.
Subsequently, when answering a query, the centralized algorithm \emph{reuses} the saved messages for $x$ and the secret vertices (without further privacy loss) when it simulates the core decomposition algorithm on the corresponding query graph. This allows it to answer many queries  correctly.

Originally, we had a more complex, direct argument. We briefly mention it here to give an idea of what is going on ``under the hood'' of the reduction. Specifically, we showed that, on a class of random graphs (similar to the query graphs instantiated in our reduction), most transcripts of a 1-round core decomposition algorithm have the property that, conditioned on seeing the transcript, the coreness of some (fixed) vertex still has high variance, $\Omega(n)$. This implies that the standard deviation of the additive error is $\Omega(\sqrt{n})$. We prefer the reduction argument, as it is simpler and gives a stronger result.

\paragraph{One-round algorithm for densest subgraph.}
In our single-round exact densest subgraph algorithm, the vertices send the server a noisy version of the incidence matrix of the graph by using randomized response. In particular, each vertex sends its row of the incidence matrix and, for each bit, it sends the actual bit with probability $p = e^{\eps}/1+e^{\eps}$; the flipped bit is sent with probability $1-p$. Using this, the server can create an unbiased estimator $\tilde{E}(S)$ for the number of edges $E(S)$ inside any \emph{fixed} subset of vertices $S$. The key observation is that the standard deviation $\sigma$ of the estimated density, i.e., $\tilde{E}(S)/|S|$, is $O(1 + \eps^{-1})$ and the estimated density is distributed binomially. Hence, by a Chernoff bound, the estimated density is $\Omega(\sqrt{\log t})$ standard deviations away from the actual density with probability at most $1/t$. By taking a union bound over all $S$, we can upper bound the total error over \emph{all} (exponentially many) subsets by $\alpha = O(\sigma\sqrt{n})$ with high probability. Hence, a \emph{brute-force}, exponential-time search, which computes the maximum $\tilde{E}(S)/|S|$ over all subsets $S$, finds a subset whose density is at most $\alpha$ from the maximum density, with high probability. The union bound appears to be necessary, but it is conceivable that there are more clever search strategies that do not require looking at exponentially many subsets $S$.

\section{Differential Privacy Background}

We give some basic definitions and background on differential privacy in this section, see~\cite{dwork2014algorithmic} for more details. Since we will only be working with graphs, we will state these results in terms of edge differential privacy for graph algorithms; we will also use the terms differential privacy and edge differential privacy interchangeably. %
Edge-neighboring graphs is a well-studied model for
differential privacy (see e.g.~\cite{li2023private} for a survey of such results) and protects the privacy of edges with highly sensitive information like
disease transmission graphs.

\begin{definition}
Two graphs $G_1=(V_1,E_1)$ and $G_2=(V_2,E_2)$ are \emph{edge-neighboring}, denoted by $G_1\sim G_2$, if $V_1=V_2$ and $|E_1\oplus E_2|=1$ (i.e., they have the same vertex set and they differ by exactly one edge).
\label{def:edge-nbhr}
\end{definition}

\begin{definition}
We use $\mathcal{G}$ to denote the set of undirected graphs. An algorithm $\mathcal{M}:\mathcal{G}\to\mathcal{Y}$ is \emph{$(\eps,\delta)$-edge differentially private} ($(\eps,\delta)$-edge dp) if for all edge-neighboring graphs $G\sim G^\prime$ and every $S\subseteq\mathcal{Y}$, we have $$\Pr[\mathcal{M}(G)\in S]\le\exp(\eps)\cdot\Pr[\mathcal{M}(G^\prime)\in S]+\delta.$$
If $\delta=0$, we say $\mathcal{M}$ is $\eps$-edge differentially private ($\eps$-edge DP).
When $\mathcal{M}$ is defined over $\mathbb{N}$ or $\mathbb{R}$ instead, it is called $\eps$-differentially private ($\eps$-DP).
\end{definition}

Next, we define two important properties of differential privacy: composition and post-processing.

\begin{lemma}
Let $\mathcal{M}_1:\mathcal{G}\to\mathcal{Y}_1$ and $\mathcal{M}_2:\mathcal{G}\to\mathcal{Y}_2$ be $\eps_1$- and $\eps_2$-edge differentially private mechanisms, respectively. Then $\mathcal{M}:\mathcal{G}\to\mathcal{Y}_1\times\mathcal{Y}_2$ defined by $\mathcal{M}=(\mathcal{M}_1,\mathcal{M}_2)$ is $(\eps_1+\eps_2)$-edge differentially private.
\label{lem:basic-comp}
\end{lemma}

\begin{lemma}
Let $\mathcal{M}:\mathcal{G}\to\mathcal{Y}$ be an $\eps$-edge differentially private mechanism.
Let $f:\mathcal{Y}\to\mathcal{Z}$ be an arbitrary randomized mapping. Then $f\circ\mathcal{M}:\mathcal{G}\to\mathcal{Z}$ is still $\eps$-edge differentially private.\label{lem:post}
\end{lemma}

The \emph{$\ell_1$-sensitivity} of a function $f : \mathcal{G} \to \mathbb{R}^d$, denoted $\Delta_1(f)$, is the supremum of the quantity $\|f(x) - f(x')\|_1 = \sum_{i=1}^d |f(x)_i - f(x')_i|$, over all neighboring $G,G' \in \mathcal{G}$. The \emph{Laplace} distribution (centered at 0) with \emph{scale} $b > 0$ has probability density function $f(x) = \frac{1}{2b} \exp(-|x|/b)$. We write $X \sim \mathsf{Lap}(b)$ or just $\mathsf{Lap}(b)$ to denote a random variable $X$ that is distributed according to the Laplace distribution with scale $b$.

\begin{fact}
\label{lem:laplace_tailbounds}
If random variable $X \sim \mathsf{Lap}(b)$, then, for any $\beta > 0$, $\Pr[|X| > b \log (1/\beta)] \le \beta$.
\end{fact}

\begin{fact}[Theorem 3.6 in \cite{dwork2014algorithmic}] \label{lem:Laplacemech}
Let $f : \mathcal{G} \to \mathbb{R}^d$ be any  function, let $\epsilon > 0$, and let, for each $i \in [d]$,  $X_i \sim \Lap(\Delta_1(f)/\epsilon)$. Then the \emph{Laplace mechanism}
$
\mathcal{A}(x)=f(x)+(X_1,\dots,X_d)
$
is $\epsilon$-differentially private.
\end{fact}

\subsection{Local Edge Differential Privacy (LEDP)}

The local model of differential privacy studies the setting where all private information is kept private by the individual parties, i.e.\ there exists no trusted curator. For graphs, the local model is called local edge differential privacy \cite{DLRSSY22}.
We now define this formally.

\begin{definition}\label{def:local-randomizer}
Let $G$ be a graph.
    An \defn{$\eps$-local randomizer} $R: \adj \rightarrow \rangeout$ for vertex $v$ is an $\eps$-edge differentially private
    algorithm that takes as input the set of its neighbors $N(v)$ in $G$, represented by
    an adjacency list $\adj = (b_1, \dots, b_{|N(v)|})$.
\end{definition}

Note that the information released via local randomizers is public to all nodes and the curator.
The curator performs some computation on the released information and makes the result public via sending a message to all vertices. The overall computation is formalized via the notion of the transcript.

\begin{definition}\label{def:LEDP}
Let $G$ be a graph.
A \emph{transcript} $\pi$ is a vector consisting of 4-tuples $(S^t_U, S^t_R, S^t_\eps, S^t_Y)$
 -- encoding the set of vertices chosen, the set of local randomizers assigned, the set of privacy parameters of the randomizers, and the set of randomized outputs produced -- for each round $t$. Let $S_\pi$ be the collection of all transcripts and $S_R$ be the collection of all randomizers. Let $\text{STOP}$ denote a special character indicating that the local randomizer's computation halts.
A \emph{protocol} is an algorithm $\alg: S_\pi \to
(2^{[n]} \times 2^{S_R} \times 2^{\mathbb{R}^{\geq 0}} \times 2^{\mathbb{R}^{\geq 0}})\; \cup \{\text{STOP}\}$
mapping transcripts to sets of vertices, randomizers, and randomizer privacy parameters. The length of the transcript, as indexed by $t$, is its round complexity. Given $\eps> 0$,  a randomized protocol $\alg$ on graph $G$ is \emph{$\eps$-local edge differentially private} if the algorithm that outputs the entire transcript generated by $\alg$ is $\eps$-edge differentially private on graph $G.$
\end{definition}

The definition is difficult to parse, but it naturally corresponds to the intuition of vertices with private information communicating with an untrusted curator (represented by the protocol) so that the curator can compute some output that depends on the private data of the vertices (see also the discussion in~\cite{DKLV23}). At the beginning, the curator only knows the vertex set $V$ and each vertex knows its incident edges, i.e., its neighborhood. In each round, a chosen  subset of the vertices performs some computation using local randomizers on (a) their local edge information, (b) the outputs from previous rounds, and (c) the public information, so basically the whole transcript so far.
The vertices then output a message which can be seen by all other vertices and the curator. The released information of each vertex is $\eps$-dp since it computed based on the output of its local randomizer. In each round the curator can choose whom to query, what local randomizer they use, and what privacy parameters the randomizer uses. Since the curator's choice only depends on the transcripts from the previous rounds, this is exactly the definition of a protocol in the above definition.

We end the discussion with a short remark on stateless/stateful vertices. As defined above, a local randomizer is only allowed to look at the past (public) transcript and its (private) neighborhood list to compute its output for the current round. This formal definition disallows the use of private states that persist across rounds in the local memory of a vertex. Private stateful behaviour is nonetheless crucial for implementing Multidimensional AboveThreshold locally, since we require each vertex to store its noisy threshold privately. This is not a problem, since one can implement a protocol as in \cite[Lemma 8]{beimel2008distributed} to simulate private states using local randomizers.
\section{Generalized Sparse Vector Technique}\label{sec:mat}

We introduce a simple, novel multidimensional generalization of the AboveThreshold mechanism, which is the basis of the algorithms in this paper. In the standard AboveThreshold mechanism from Section 3.6 in~\cite{dwork2014algorithmic}, we are given as input a threshold $T$ and an adaptive stream of sensitivity $1$ queries $f_1,f_2,\ldots$, and the goal is to output the first query which exceeds the threshold. In the multidimensional version, we have a $d$-dimensional vector of thresholds $\vec{T}=(T_1,\ldots,T_d)$ and an adaptive stream of $\ell_1$-sensitivity $D$ vector-valued queries $\vec{f}_1,\vec{f}_2,\ldots$. The goal is to output, for each coordinate $j\in[d]$, the first query $i_j$ for which the $j^{th}$ coordinate exceeds the threshold $T_j$. See \Cref{alg:multidimensional Above Threshold} for the pseudo-code of the algorithm.

Here $D$ is defined as the sensitivity for \emph{all} dimensions, namely, for edge-neighboring graphs $G \sim G'$, %
\begin{align}
D := \max_{G \sim G'} \left(
\sum_{j \in [d]} \left(\max_{i \in [I_j]}\left(|f_{i,j}(G) - f_{i,j}(G')|\right)\right)\right),
\end{align}
where $I_j$ is the number of queries in dimension $j$.

\begin{algorithm}[t]
\caption{Multidimensional AboveThreshold (MAT)}
\label{alg:multidimensional Above Threshold}
\textbf{Input:} Private graph $G$, adaptive queries $\{\vec{f}_1, \dots, \vec{f}_n\}$, threshold vector $\vec{T}$, privacy $\eps$,
sensitivity $D$.\\
\textbf{Output:} A sequence of responses $\{\vec{a}_1, \dots, \vec{a}_n\}$ where $a_{i,j}$ indicates if $f_{i,j}(G)\ge \vec{T}_j$\\
\begin{algorithmic}[1]
\FOR{$j=1,\ldots,d$}
    \STATE $\hat{T}_j\leftarrow \vec{T}_j+\text{Lap}(2D/\eps)$
\ENDFOR
\STATE
\FOR{each query $\vec{f}_i \in \{\vec{f}_1, \dots, \vec{f}_n\}$}
\FOR{$j=1,\ldots,d$}
\STATE Let $\nu_{i,j}\leftarrow\text{Lap}(4D/\eps)$
\IF{$f_{i,j}(G)+\nu_{i,j}\ge \hat{T}_j$}
\STATE Output $a_{i,j}=\top$
\STATE Stop answering queries for coordinate $j$
\ELSE
\STATE Output $a_{i,j}=\bot$
\ENDIF
\ENDFOR
\ENDFOR
\end{algorithmic}
\end{algorithm}

\begin{theorem}
    \Cref{alg:multidimensional Above Threshold} is $\eps$-differentially private.\label{thm:privacy AboveThreshold}
\end{theorem}
\begin{proof}
Fix $G\sim G^\prime$ to be arbitrary edge-neighboring graphs.
Let $\mathcal{A}(G)$ and $\mathcal{A}(G')$ denote the random variable representing the output of \Cref{alg:multidimensional Above Threshold} when run on $G$ and $G'$, respectively. For each coordinate $j\in [d]$, there is some $r(j)\in\{1,\ldots,n+1\}$ such that the output satisfies $a_{\cdot j}\in\{\top,\bot\}^{r(j)}$ and has the form that for all $i<r(j)$, $a_{i,j}=\bot$. We can assume without loss of generality that $a_{r(j),j}=\top$; if $r(j)<n+1$, this must be the case and if $r(j)=n+1$, we can simply ask an additional query which evaluates to $\infty$, independent of the dataset.
Our goal is to show that
for any output $a$, we have
$$\Pr[\mathcal{A}(G)=a]\le \exp(\eps)\cdot\Pr[\mathcal{A}(G')=a].$$
Observe that there are two sources of randomness in the algorithm: the noisy thresholds $\hat{T}_j$ for each $j\in[d]$ and the perturbations of the queries $\nu_{i,j}$ for each $j\in[d]$, $i\in[r(j)]$. We fix the random variables $\nu_{i,j}$ for each $j\in[d]$ and $i<r(j)$ by simply conditioning on their randomness; for neighboring graphs, we have a coupling between corresponding variables.
For simplicity, we omit this from notation. The remaining random variables are $\hat{T}_j$ and $\nu_{r(j),j}$ for each $j\in [d]$; we will be taking probabilities over their randomness.
The main observation needed is that the output of $\mathcal{A}$ is uniquely determined by the values of  $r(j)$ for each $j\in [d]$. Thus, we can define the (deterministic, due to conditioning on the probabilities) quantity
$$g_j(G)\coloneqq \text{max}\{f_{i,j}(G)+\nu_{i,j}:i<r(j)\}$$
for each $j\in[d]$ so that the event that $\mathcal{A}(G)=a$ is equivalent to the event where
\begin{align}
 \hat{T}_j > g_j(G)
 \text{ and }
    f_{r(j),j}(G)+\nu_{r(j),j} \geq \hat{T}_j\text{ for each } j\in[d].\label{eq:event}
\end{align}%
Let $p(\cdot)$ and $q(\cdot)$ denote the density functions of the random vectors consisting of $\hat{T}(j)$ and $\nu_{r(j),j}$ for each $j\in[d]$. Thus,
\begin{align}
    \Pr[\mathcal{A}(G)=a]&=\Pr[\hat{T}_j\in(f_{r(j),j}(G)+\nu_{r(j),j},g_j(G))\text{ $\forall j\in [d]$}]\nonumber
\end{align}
Using $v_j'$ to denote the value of $\nu_{r(j), j}$ and $t_j'$ to denote the value of $\hat{T}_j$,
\begin{align}
    \Pr[\mathcal A(G) = a]
    &=\int_{\mathbb{R}^d}\int_{\mathbb{R}^d}p(\vec{t'})q(\vec{v'})\cdot\mathbbm{1}\{t_j'\in(f_{r(j),j}(G)+v_j',g_j(G))\text{ $\forall j\in [d]$}\}\,d\vec{v'}\,d\vec{t'}\label{eq:star}
\end{align}
Now, we make a change of variables from $\vec{v'}$ to $\vec{v}$ and $\vec{t'}$ to $\vec{t}$, with the change given by
\begin{align*}
    {v}'_j&={v}_j+g_j(G)-g_j(G^\prime)+f_{r(j),j}(G')-f_{r(j),j}(G)\\
    {t}'_j&={t}_j+g_j(G)-g_j(G^\prime)
\end{align*}
for each $j\in [d]$.
We have that $d\vec{t'}_j = d\vec{t}_j$, and $d\vec{v'}_j = d\vec{v}_j$ since the new variables differ from the old ones by a constant that is independent of $\vec{v}$ and $\vec{t}$. The term inside the indicator function in the integral now becomes
\[
t_j + g_j(G) - g_j(G') \in
( f_{r(j), j} (G) + v_j + g_j(G) - g_j(G') + f_{r(j), j}(G') - f_{r(j), j}(G), g_j(G))
\forall j \in [d]
\] which on subtracting $g_j(G) - g_j(G')$ gives
\[
t_j \in ( f_{r(j), j} (G') + v_j , g_j(G')) \forall j \in [d]
\]
Thus, we can apply the change of variables to rewrite (\ref{eq:star}) as
\begin{align}
    \int_{\mathbb{R}^d}\int_{\mathbb{R}^d}p(\vec{t'})q(\vec{v'})\mathbbm{1}\{{t}_j\in(f_{r(j),j}(G^\prime)+{v}_j,g_j(G^\prime))\text{ $\forall j\in [d]$}\}\,d\vec{v}\,d\vec{t}.\nonumber
\end{align}

Since we have conditioned on the randomness of $\nu_{i,j}$ for each $j\in[d]$ and $i<r(j)$, we have $\|\vec{v'}-\vec{v}\|_1\le 2D$  and $\|\vec{t'}-\vec{t}\|_1\le D$ since the vectors $f_{i\cdot}(G)$ have sensitivity $D$, implying that the vectors $g_{\cdot}(G)$ also have sensitivity $D$. %
This implies that $p(\vec{t}^\prime)\le\exp(\eps/2)\cdot p(\vec{t})$ and $q(\vec{v}^\prime)\le\exp(\eps/2)\cdot q(\vec{v})$ by the properties of the Laplace distribution. We can thus upper bound the above integral by
\begin{align}
    \exp(\eps)\cdot\int_{\mathbb{R}^d}\int_{\mathbb{R}^d}p(\vec{t})q(\vec{v})\mathbbm{1}\{t_j\in(f_{r(j),j}(G^\prime)+{v}_j,g_j(G^\prime))\text{ $\forall j\in [d]$}\}\,d\vec{v}\,d\vec{t}\nonumber
\end{align}
Rewriting the integral as the probability of an event, we get
\begin{align*}
    \exp(\eps)\cdot\textstyle\Pr[\hat{T}_j\in(f_{r(j),j}(G^\prime)+\nu_{r(j),j},g_j(G^\prime))\text{ $\forall j\in [d]$}].
\end{align*}
Finally, by our observation in (\ref{eq:event}), this is equal to
\begin{align*}
    \exp(\eps)\cdot\Pr[\mathcal{A}(G')=a].
\end{align*}
Putting the inequalities together, we get
$$\Pr[\mathcal{A}(G)=a]\le \exp(\eps)\cdot\Pr[\mathcal{A}(G')=a].$$
Since $G,G^\prime$ were arbitrary edge-neighboring graphs, this completes the proof.
\end{proof}

\subsection{LEDP Multidimensional AboveThreshold (MAT) Mechanism}

In our algorithms, we apply the multidimensional AboveThreshold mechanism in the special case where the queries $\vec{f}_i$ have $d=n$ coordinates, and each coordinate of the query corresponds to a node $u$ in the graph $G$ and only depends on the edges $e=(u,v)$ for $v\in V-\{u\}$. In other words, we have an instance of the AboveThreshold on each node $u\in V$, where the data used for the AboveThreshold instance is only the local (edge) data of $u$.
We will show that in this setting, the multidimensional AboveThreshold mechanism can be implemented locally to satisfy local edge-differential privacy.

For clarity of notation, we index the coordinates of the queries and threshold vector by nodes $v\in V$ instead of indices $j\in[n]$. That is, each query consists of $\vec{f}_{i,v}$ for $v\in V$ and the threshold vector consists of $\vec{T}_{v}$ for $v\in V$. We will now present the changes needed for the local implementation. In Lines 1--3, let each node $u\in V$ compute and store its noisy threshold $\hat{T}_u$ using the public threshold $\vec{T}_u$. Then for each query $\vec{f}_i$ in lines 5--15, each node $u\in V$ can sample its own noise $\nu_{i,u}$ and check the condition $f_{i,u}(G)+\nu_{i,u}\ge\hat{T}_u$. The pseudocode is given in \Cref{alg:local multidimensional Above Threshold}, where actions performed by Node $v$ in the algorithm are performed their corresponding local randomizers, and we now show that this is locally edge-differentially private.

\begin{algorithm}[t]
\caption{Local Multidimensional AboveThreshold}
\label{alg:local multidimensional Above Threshold}
\textbf{Input:} Private graph $G$, adaptive queries $\{\vec{f}_i\}$, threshold vector $\vec{T}$, privacy $\eps$,
sensitivity $D$.\\
\textbf{Output:} A sequence of responses $\{\vec{a}_i\}$ where $a_{i,j}$ indicates if $f_{i,j}(G)\ge \vec{T}_j$\\
\begin{algorithmic}[1]
\FOR{$v\in V$}
    \STATE Node $v$ computes $\hat{T}_v\leftarrow \vec{T}_v+\text{Lap}(2D/\eps)$ and stores it
\ENDFOR
\STATE
\FOR{each query $\vec{f}_i$}
\FOR{$v\in V$}
\STATE Node $v$ samples $\nu_{i,v}\leftarrow\text{Lap}(4D/\eps)$
\IF{$f_{i,v}(G)+\nu_{i,v}\ge \hat{T}_v$}
\STATE Node $v$ outputs $a_{i,j}=\top$
\STATE Node $v$ outputs $\text{STOP}$, and stops answering queries
\ELSE
\STATE Node $v$ outputs $a_{i,v}=\bot$
\ENDIF
\ENDFOR
\ENDFOR
\end{algorithmic}
\end{algorithm}

\begin{theorem}\label{thm:mat-ledp}
    \Cref{alg:local multidimensional Above Threshold} is $\eps$-locally edge differentially private.
\end{theorem}
\begin{proof}
    First, we need to show that the local randomizers are in fact edge-differentially private. This can be seen because each randomizers' output is a subset of the output of \Cref{alg:multidimensional Above Threshold} and it is only computed based on the incident edges of each vertex. In other words, the output is a post-processing of \Cref{alg:multidimensional Above Threshold}, so the privacy guarantees follow from \Cref{thm:privacy AboveThreshold} and \Cref{lem:post}. Next, we argue that the transcript is also $\eps$-edge differentially private. Recall that the transcript consists of the set of vertices chosen, the set of local randomizers assigned, the set of privacy parameters assigned, and the set of outputs. In the algorithm, the set of vertices chosen, the set of local randomizers assigned, and the set of privacy parameters assigned are all functions of the outputs from the previous rounds and the public information. Thus, it suffices to prove that the outputs are differentially private by post-processing (\Cref{lem:post}). But this was already proven in \Cref{thm:privacy AboveThreshold}, so we are done.
\end{proof}
\section{\texorpdfstring{LEDP $k$-Core Decomposition}{Differentially Private k-Core Decomposition}}

In this section, we give our improved algorithm for differentially private $k$-core decomposition. We first present a variant of the classical (non-private) algorithm for $k$-core decomposition in \Cref{sec:optimal-variant}, and then show how to make it private in \Cref{sec:optimal-private}. Finally, we show how to implement it in near-linear time in \Cref{sec:optimal-efficient}, while incurring only $(1+\eta)$-multiplicative error.

\subsection{A Variation of the Classical Algorithm}
\label{sec:optimal-variant}

The classical peeling algorithm begins with the full vertex set $V$, which is the $0$-core of the graph. Given the $(k-1)$-core, the algorithm computes the $k$-core via an iterative peeling process: the algorithm repeatedly removes all nodes $v$ for which the induced degree is less than $k$, and labels the nodes which remain as being part of the $k$-core. Running this for $k$ from 1 up to $n$ gives the full algorithm, the pseudocode of which is given in \Cref{alg:classical-algorithm}.

\begin{algorithm}[ht]
\caption{Threshold-Based $k$-core Decomposition Algorithm}
\label{alg:classical-algorithm}
\textbf{Input:} Graph $G=(V,E)$.\\
\textbf{Output:} $k$-core value of each node $v\in V$\\
\begin{algorithmic}[1]
\STATE Initialize $V_0\leftarrow V$, $t\leftarrow 0$, $\hat{k}(v)\leftarrow 0$ for all $v\in V$ 
\FOR{$k=1,\ldots,n$}\label{line:for-k}
\REPEAT
\STATE $t\leftarrow t+1$, $V_t\leftarrow V_{t-1}$
\FOR{$v\in V_{t-1}$}
\IF{$d_{V_{t-1}}(v)< k$}\label{line:classical if condition}
\STATE $V_t\leftarrow V_t-\{v\}$\label{line:classical peeled}
\ENDIF{}
\ENDFOR{}
\UNTIL{$V_{t-1}-V_{t} = \emptyset$}
\STATE
\STATE Update the core numbers $\hat{k}(v)\leftarrow k$ for all nodes $v\in V_t$
\ENDFOR
\end{algorithmic}
\end{algorithm}

\begin{theorem}
    For each $v\in V$, the output $\hat{k}(v)$ given by \Cref{alg:classical-algorithm} is the core number of $v$.
\end{theorem}
\begin{proof}
    We will inductively show that the algorithm recovers the $k$-core of the graph. The base case of $k=0$ is easy. Now, assume that the algorithm finds the true $(k-1)$-core. Let $V(k)$ denote the subset of nodes which aren't removed in the iterative process for $k - 1$ in Line~\ref{line:for-k}. We have that each node $v\in V(k)$ has induced degree at least $k$ in $V(k)$, or else it would have been removed. Thus, we know that the core numbers $k(v)$ is at least $k$ for each $v\in V(k)$, so we have that $V(k)$ is a subset of the true $k$-core. Now, let $K$ denote the true $k$-core. Since the $k$-core is always a subset of the $(k-1)$-core by definition, each node $v\in K$ is in $V_t$ at the beginning of the iterations. Furthermore, we know that $v\in K$ is never removed from $V_t$ since the induced degree is always at least $k$ since $K\subseteq V_t$. Thus, we have that $V(k)$ is a superset of the true $k$-core as well. Thus, $V(k)$ is the true $k$-core for each $k$, so all nodes are labelled correctly.  
\end{proof}

\subsection{Private Implementation of the Algorithm}
\label{sec:optimal-private}

It is difficult to turn the classical algorithm into a differentially private one because it has $\Omega(n)$ iterations, which causes us to incur $\tilde{\Omega}(n)$ additive error when using basic composition (\Cref{lem:basic-comp}). In fact, this was cited  in~\cite{DLRSSY22} as the motivation for basing their algorithms on parallel/distributed algorithms for $k$-core decomposition, since those algorithms often have $\text{poly}\log({n})$ round-complexity. Our main observation is that the private version of \Cref{alg:classical-algorithm}     can be analyzed as a special case of the Multidimensional AboveThreshold mechanism, so it doesn't need to incur the $\Omega(n)$ additive error due to composition. 

\begin{algorithm}[ht]
\caption{$\eps$-Differentially Private $k$-Core Decomposition}
\label{alg:optimal-algorithm}
\textbf{Input:} Graph $G=(V,E)$, privacy parameter $\eps>0$.\\
\textbf{Output:} An $(1,120\log(n)/\eps)$-approximate $k$-core value of each node $v\in V$\\
\begin{algorithmic}[1]
\STATE $V_0\leftarrow V$, $t\leftarrow 0$, $k=60\log{n}/\eps$.
\STATE Initialize $\hat{k}(v)\leftarrow 0$ for all $v\in V$ \label{line: initial labelling}
\FOR{$v\in V$}
\STATE $\tilde{\ell}(v)\leftarrow \text{Lap}(4/\eps)$
\ENDFOR{}
\STATE
\WHILE{$k\le n$}\label{line:optimal k-iteration}
\REPEAT
\STATE $t\leftarrow t+1$, $V_t\leftarrow V_{t-1}$
\FOR{$v\in V_{t-1}$}
\STATE $\nu_{t,v}\leftarrow\text{Lap}(8/\eps)$
\IF{$d_{V_{t-1}}(v)+\nu_{t,v}\le k+\tilde{\ell}(v)$}\label{line:optimal if condition}
\STATE $V_t\leftarrow V_t-\{v\}$\label{line:optimal peeled}
\ENDIF{}
\ENDFOR{}
\UNTIL{$V_{t-1}-V_{t} = \emptyset$}
\STATE
\STATE Update the core numbers $\hat{k}(v)\leftarrow k$ for all nodes $v\in V_t$
\STATE $k\leftarrow k+60\log{n}/\eps$
\ENDWHILE{}
\end{algorithmic}
\end{algorithm}

\begin{theorem}
    \Cref{alg:optimal-algorithm} is $\eps$-edge differentially private.\label{thm:optimal-privacy}
\end{theorem}
\begin{proof}
    We will show that \Cref{alg:optimal-algorithm} is an instance of the multidimensional AboveThreshold mechanism, implying that it is $\eps$-edge differentially private. Specifically, we will show that its output can be obtained by post-processing the output of an instance of \Cref{alg:multidimensional Above Threshold}. Indeed, consider the instance of the multidimensional AboveThreshold mechanism where the input graph is $G$, the privacy parameter is $\eps$, and the threshold vector $\vec{T}=\vec{0}$ is the zero vector. We will now inductively (and adaptively) define the queries. 
    
    For each iteration $t$, the $t^{th}$ query consists of the vector of $k-d_{V_{t-1}}(v)$ for each $v\in V$, where $V_0=V$ as in the algorithm; note that this matches with the queries on Line \ref{line:optimal if condition} of the algorithm. First, observe that the sensitivity of the vector queries of $k-d_{V_{t-1}}(v)$ for each $v\in V$ is $D=2$ since one edge change will only affect two coordinates of the query (each by 1), as needed in the privacy analysis. Next, observe that we can construct $V_t$ from $V_{t-1}$ using only the outputs $a_t(v)$ of the queries at the current iteration: given $v\in V_{t-1}$, we include $v$ in $V_t$ as well if and only if $a_t(v)=\bot$, which is equivalent to the condition in Line \ref{line:optimal if condition} of $d_{V_{t-1}}(v)+\nu_{t,v}\le k+\tilde{\ell}(v)$\footnote{Here, we implicitly assume that the randomness used by the algorithm and the AboveThreshold mechanism are the same. This is justified by a coupling of the random variables in the two algorithms. Specifically, we couple the noise added in Line 4 of the algorithm with the noise added in Line 2 of the AboveThreshold mechanism and couple the noise $\nu_{t,v}$ with $\nu_{i,j}$ in the AboveThreshold mechanism. It is easy to see that for $\Delta=2$, the random variables are exactly the same so the privacy guarantees of multidimensional AboveThreshold translates to privacy guarantees for our algorithm.}. Thus, this is a feasible sequence of adaptive queries for the AboveThreshold mechanism.
    
    Since the sequence of subsets $\{V_t\}$ are obtained by post-processing the outputs $a_t(v)$ of the AboveThreshold mechanism, they are $\eps$-differentially private by \Cref{thm:privacy AboveThreshold} and \Cref{lem:post}. By applying post-processing again, the sequence of pairs $(V_t,k_t)$ for each iteration $t$ is also $\eps$-differentially private since the sequence $k_t$ is public. Given the pairs $(V_t,k_t)$ for each iteration $t$, we can now recover the approximate core numbers which the algorithm outputs by setting the core numbers as $\hat{k}(v)=k_t$ for each node in $V_t-V_{t-1}$. It is easily verified that this gives the exact same output as \Cref{alg:optimal-algorithm}, so we have $\eps$-differential privacy for the algorithm by applying post-processing (\Cref{lem:post}) once again.
\end{proof}

\begin{theorem}
    \Cref{alg:optimal-algorithm} outputs $(1,\frac{120\log{n}}{\eps})$-approximate core numbers $\hat{k}(v)$ with probability $1-O(\frac{1}{n^2})$.\label{thm:optimal-utility}
\end{theorem}
\begin{proof}
By the density function of the Laplace distribution, we know that we have $|\nu_{t,u}|\le \frac{40\log{n}}{\eps}$ and $|\tilde{\ell}(u)|\le \frac{20\log{n}}{\eps}$ each with probability at least $1-\frac{1}{n^5}$. Since there are at most $O(n^3)$ such random variables, taking a union bound over all nodes $u\in V$ and all iterations of the loops, we have the above guarantee with probability at least $1-{O}\left(\frac{1}{n^2}\right)$. We condition on the event that the above inequalities hold true for each $t\in[T]$ and each $u\in V$ for the remainder of the proof.

Fix an arbitrary iteration $k$ of the while loop starting on Line \ref{line:optimal k-iteration}. 
Let $H$ be the set of nodes remaining in $V_t$ at the end of the while loop in Lines 8---16. We claim that ($i$) all nodes $u\in H$ have core number at least $k-\frac{60\log{n}}{\eps}$ and ($ii$) all nodes $u\not\in H$ have core number at most $k+\frac{60\log{n}}{\eps}$. To see ($i$), consider the subgraph $H$; we claim that each node $u\in H$ has induced degree at least $k-\frac{60\log{n}}{\eps}$ in $H$. Indeed, since each node in $H$ was not removed in the final iteration of the while loop in Lines 8---16, we have that $d_{H_t}(u)+\nu_{t,u}\ge k+\tilde{\ell}(u)$ for each $u\in H$. But since we have assumed $|\nu_{t,u}|\le\frac{40\log{n}}{\eps}$ and $|\tilde{\ell}(u)|\le\frac{20\log{n}}{\eps}$, our desired bounds follow directly by the triangle inequality. To see ($ii$), let's suppose for contradiction that the $k$-core value of $u$ is $k(u)>\ell+\frac{60\log{n}}{\eps^\prime}$. Then there exists a subgraph $u\in K\subseteq V$ where the induced degree of each node $v\in K$ is $d_K(v)\ge k(u)$. But for such a subgraph $K$, the condition in Line 12 will always be true (again, because $|\nu_{t,u}|\le\frac{40\log{n}}{\eps}$ and $|\tilde{\ell}(u)|\le\frac{20\log{n}}{\eps}$) so $K\subseteq H$. But since $u\in K$, this contradicts the fact that $u\not\in H$.

Using the above bounds, we can now prove our desired results. First, consider an arbitrary node $u\in V$ labeled $\hat{k}(u)$ within the while loop in Lines \ref{line:optimal k-iteration}---20 and not relabeled later. Since $u$ was labeled $\hat{k}(u)$ in the current iteration, we have that $k(u)\ge \hat{k}(u)-\frac{60\log{n}}{\eps^\prime}$. Since $u$ was not relabelled in the next iteration where the threshold was $k=\hat{k}(u)+\frac{60\log{n}}{\eps}$, the node $u$ was removed from $V_t$ at that iteration. Consequently, we have $k(u)\le k+\frac{60\log{n}}{\eps}=\hat{k}(u)+\frac{120\log{n}}{\eps}$ by what we just proved above. Since the output of our algorithm is $\hat{k}(u)$, the desired bounds in the theorem statement follow directly. Now, let's consider an arbitrary node $u\in V$ labeled $0$ at Line \ref{line: initial labelling} at the beginning of the algorithm and not relabeled later. Since the node $u$ was not relabelled at the first iteration where $k=\frac{60\log{n}}{\eps}$, it was removed from $V_t$ at that iteration, implying that $k(u)\le \frac{120\log{n}}{\eps}$. Hence, we have the desired approximation guarantees for all nodes.
\end{proof}

The above proof can be modified with appropriate constants to give the approximation with probability 
at least $1 - \frac{1}{n^c}$ for any $c \geq 1$, thus yielding our desired with high probability result.

\subsection{Efficient Implementation of the Algorithm}
\label{sec:optimal-efficient}

Though the current algorithm has strong utility guarantees, the running time is at least quadratic in $n$. We will make a slight modification to the algorithm, which will incur an additional multiplicative error of $1+\eta$ for an user chosen constant $\eta>0$, and show that a clever sampling trick can be used to implement the resulting algorithm in time nearly linear in the number of edges $m$. The only difference in \cref{alg:efficient-algorithm} and \cref{alg:optimal-algorithm} is in Line 19, where the threshold is increased by a multiplicative $1+\eta$ instead of an additive $60\log{n}/\eps$. Via a similar proof as before, we can obtain an analogous set of privacy and utility guarantees.

\begin{lemma}\label{thm:efficient-1}
    \Cref{alg:efficient-algorithm} has the following guarantees:
    \begin{itemize}
        \item It is $\eps$-edge differentially private.
        \item It outputs $(1+\eta,O(\log{n}/\eps))$-approximate core numbers with probability $1-O(\frac{1}{n^2})$.
    \end{itemize}
\end{lemma}

\begin{algorithm}[ht]
\caption{}
\label{alg:efficient-algorithm}
\textbf{Input:} Graph $G=(V,E)$, multiplicative error $\eta>0$, privacy parameter $\eps>0$.\\
\textbf{Output:} An $(1+\eta,60\log(n)/\eps)$-approximate $k$-core value of each node $v\in V$\\
\begin{algorithmic}[1]
\STATE $V_0\leftarrow V$, $t\leftarrow 0$, $k=60\log{n}/\eps$.
\STATE Initialize $\hat{k}(v)\leftarrow 0$ for all $v\in V$ \label{line: efficient initial labelling}
\FOR{$v\in V$}
\STATE $\tilde{\ell}(v)\leftarrow \text{Lap}(4/\eps)$
\ENDFOR{}
\STATE
\WHILE{$k\le n$}\label{line:efficient k-iteration}
\REPEAT
\STATE $t\leftarrow t+1$, $V_t\leftarrow V_{t-1}$
\FOR{$v\in V_{t-1}$}
\STATE $\nu_{t,v}\leftarrow\text{Lap}(8/\eps)$
\IF{$d_{V_{t-1}}(v)+\nu_{t,v}\le k+\tilde{\ell}(v)$}\label{line:efficient if condition}
\STATE $V_t\leftarrow V_t-\{v\}$\label{line:efficient peeled}
\ENDIF{}
\ENDFOR{}
\UNTIL{$V_{t-1}-V_{t} = \emptyset$}
\STATE
\STATE Update the core numbers $\hat{k}(v)\leftarrow k$ for all nodes $v\in V_t$
\STATE $k\leftarrow (1+\eta)\cdot k$
\ENDWHILE{}
\end{algorithmic}
\end{algorithm}

Now, we discuss the efficient sampling procedure. We claim that the naive implementation of the algorithm given in \cref{alg:efficient-algorithm} requires $\Omega(n^2)$ runtime in the worst case. Let us focus on the do-while loop in Lines 8--16 of \cref{alg:efficient-algorithm}. Clearly, the for-loop in Lines 10--15 iterates $\Theta(n)$ times for each iteration of the do-while loop. We claim that in the worst case, the do-while loop also requires $\Omega(n)$ iterations. Let us illustrate in the easy case where there is no noise (or equivalently, when we take $\eps$ to infinity). Consider the path graph, and observe that the degree of each node is $2$ except the endpoints which have degree $1$. If the threshold $\ell$ is 2, then only the endpoints are removed and the resulting graph will again be a path graph, now of size $n-2$. Clearly, it will take $n/2$ iterations before the do-while loop terminates. This example can be generalized to the noisy version by increasing the separation of the degrees from $1$ up to $4\log{n}$.

We will give a more efficient way to implement the sampling so that the algorithm can run in $\tilde{O}(m+n)$ time. The idea is as follows: given a vertex $v$ and a fixed threshold $k+\tilde{\ell}$, we can calculate the probability $p$ that the vertex is removed at a given iteration of the do-while loop. In fact, this probability remains the same for vertex $v$ unless a neighbor of $v$ is removed from $H_t$. Thus, whenever a neighbor of $v$ is removed, we can sample a (geometric) random variable for when the vertex $v$ is peeled assuming no more neighbors of $v$ are removed. Whenever another neighbor of $v$ is removed, we have to resample this geometric random variable, but the number of times this resampling must be done over all vertices $v$ scales linearly with the number of edges $m$ rather than quadratically with the number of nodes $n$. This is detailed below. In~\cref{alg:algorithm-efficient}, $\text{Geom}(q)$ is the \emph{geometric} distribution with parameter $q$.

\begin{algorithm}[ht]
\caption{An efficient implementation of Lines 8--16 in \cref{alg:efficient-algorithm}}
\label{alg:algorithm-efficient}
\hspace*{\algorithmicindent} \textbf{Input:} Graph $G=(V,E)$, parameter $\eps>0$, threshold $k+\tilde{\ell}(v)$ for each $v\in V$.\\
\hspace*{\algorithmicindent} \textbf{Output:} $V_t$ with same distribution as Lines 8--16 of \cref{alg:efficient-algorithm}\\
\begin{algorithmic}[1]
\STATE $t\leftarrow0$, $V_0\leftarrow V$
\STATE $\text{to-remove}(t)\leftarrow\emptyset$ for each $t\in [n]\hfill$\textcolor{Green}{//set of nodes to remove from $V_{t-1}$ at time $t$}
\STATE $\text{remove-time}(v)\leftarrow n$ for each $v\in V$
\STATE $\text{updated}\leftarrow V$
\REPEAT
\FOR{$v\in\text{updated}$}
\STATE remove node $v$ from to-remove(remove-time$(v))$\hfill\textcolor{Green}{//updates the sampled time for removing $v$}
\STATE $q\leftarrow \Pr[d_{V_t}(v)+\text{Lap}(8/\eps)\ge k+\tilde{\ell}(v)]$
\STATE $\text{remove-time}(v)\leftarrow t+\text{Geom}(q)$
\STATE add node $v$ to to-remove(remove-time$(v))$
\ENDFOR
\STATE
\STATE $t\leftarrow t+1$, $V_t\leftarrow V_{t-1}$\hfill\textcolor{Green}{//this is only notational; we only store the set $V_t$ for a single $t$}
\STATE
\FOR{$v\in\text{to-remove}(t)$}\label{line:to-remove}
\STATE remove $v$ from $V_t$
\STATE add all neighbors of $v$ in $V_t$ to \text{updated}\hfill\textcolor{Green}{//need to update remove-time if degree changes}
\ENDFOR{}
\UNTIL{$V_{t-1}-V_{t} = \emptyset$}
\end{algorithmic}
\end{algorithm}

\begin{lemma}
    \cref{alg:algorithm-efficient} satisfies the following:
    \begin{itemize}
        \item It gives the same distribution of $V_t$ as Lines 8--16 in \cref{alg:efficient-algorithm}.
        \item It can be implemented in $\tilde{O}(n+m)$ time.
    \end{itemize}\label{lem:run time}
\end{lemma}
\begin{proof}
To see that the distribution is the same, we use induction on $t$. We claim that at each iteration $t$, the distribution of $V_t$ is the same for the two implementations of the algorithm. The statement is clear for $t=0$ since $V_0=V$ in both implementations. Now, assume that the distribution of $V_{t-1}$ is the same for both implementations; we will show that the distribution of $V_{t}$ conditioned on $V_{t-1}$ is also the same for both implementations. This will imply our desired result by induction.

Given $V_{t-1}$, let us compute that the probability that a node $v\in V_{t-1}$ is not included in $V_t$ in both implementations. Note that the events that each node is not included in $V_t$ are independent (conditioned on $V_{t-1}$), so showing that the probabilities are the same is sufficient to show the equivalence of (conditional) distributions. In the implementation in \cref{alg:efficient-algorithm}, the probability is clearly just
\begin{align}
    \Pr[d_{V_{t-1}}(v)+\text{Lap}(8/\eps)\ge k+\tilde{\ell}(v)].\label{eq:prob-remove}
\end{align}
Now, let's consider the implementation in \cref{alg:algorithm-efficient}. To understand when the node $v$ will be removed, we look at the final time $t^\prime$ that node $v$ is added to the set \textit{updated}. The node $v$ will be removed at time $t$ if the \textit{remove-time(v)} sampled by the algorithm is exactly $t$. The probability that this occurs conditioned on the event that $v$ was not removed at $V_{t-1}$ can be calculated as:
$$\Pr[t^\prime+\text{Geom}(q)=t \mid t^\prime+\text{Geom}(q)\ge t-1]=\Pr[\text{Geom}(q)=1]=q,$$
where the first equality holds by the memoryless property of the geometric distribution. Since no neighbors of $v$ were removed between iterations $t^\prime$ and $t$, the induced degrees $d_{V_{t^\prime}}(v)$ and $d_{V_t}(v)$ are the same, which implies that
$q = \Pr[d_{V_{t}}(v)+\text{Lap}(8/\eps)\ge k+\tilde{\ell}(v)]$. It is thus evident that the probability $q$ is the same as the expression in \cref{eq:prob-remove}, proving the desired claim.

Now, we will prove that the implementation given in \cref{alg:algorithm-efficient} runs in near-linear time. First, we note that all sets are implemented as binary search trees so adding and removing elements requires only $O(\log{n})$ time. To analyze the time complexity, observe that the total running time of Lines 6--11 (over all iterations of the while-loop) scales with the total number of nodes added to the set \textit{updated} (up to logarithmic factors). Since a node is only added to the set \textit{updated} when one of its neighbors is removed, the total number of nodes added to \textit{updated} during the entire algorithm scales with the number of edges. The total time complexity (over all executions of the repeat loop) of Lines 16 is $\tilde{O}(n)$ since each node is only removed once and the total time complexity of Line 17 is $\tilde{O}(m)$ for the same reason as before. That completes the proof.
\end{proof}

Combining \Cref{thm:efficient-1} and \Cref{lem:run time}, we have the following theorem.

\begin{theorem}\label{thm:efficient-final}
    \Cref{alg:efficient-algorithm} has the following guarantees:
    \begin{itemize}
        \item It is $\eps$-edge differentially private.
        \item It outputs $(1+\eta,O(\log{n}/\eps))$-approximate core numbers with probability $1-O(\frac{1}{n^2})$.
        \item It can be implemented in $\tilde{O}(n+m)$ time.
    \end{itemize}
\end{theorem}

\subsection{\texorpdfstring{An Improved Analysis of \cite{DLRSSY22}}{An Improved Analysis of DLR+22}}\label{sec:improved-distributed}

In this section we use our MAT technique above to improve the algorithm of~\cite{DLRSSY22} so that we obtain an $\eps$-LEDP algorithm for
\kc decomposition with additive error $O\left(\frac{\log n}{\eps}\right)$ while maintaining the original
$O(\log^2 n)$ round complexity.  We reframe Algorithm 3 of~\cite{DLRSSY22}
within our MAT framework in the pseudocode given in~\cref{alg:improved-ledp-kcore}.

Our algorithm works as follows. The algorithm assigns \emph{levels} to each vertex representing roughly the current estimated
core number for that vertex. Initially, all levels are set to $0$ for all $L_r[i]$
(Line~\ref{lkcsr-line:initialize-L}).
We iterate through $r = 4\log^2 n$ rounds (Line~\ref{lkcsr-line:iterate}) and all $i \in [n]$ vertices (Line~\ref{lkcsr-line:iterate-vertex})
where in each round $r$ and for each vertex $i$, we store the
current \emph{level} of vertex $i$ in $L_r$ (Line~\ref{lkcsr-line:old-level}).
In each round $r$, we determine whether a vertex $i$ moves up a level by first
checking that $i$ is on the same level as the current round number (Line~\ref{lkcsr-line:old-level}).
If it is, then we perform a MAT query to determine whether the
number of neighbors (in the adjacency list $\adj_i$ of $i$ in Line~\ref{lkcsr-line:compute-up})
that is on level $r$ exceeds the threshold $\upexp^{\lfloor r/(\numgrouplevels)\rfloor}$
(Line~\ref{lkcsr-line:move-up-condition}). Noises for the MAT query are picked in Line~\ref{lkcsr-line:noise-1} and Line~\ref{lkcsr-line:sample-noise}.
If the query falls below the threshold, then we set a local variable $A_i \leftarrow 0$ and also stop performing
queries for $i$ (meaning no future queries will be made for $i$) in Lines~\ref{lkcsr-line:move-up-condition} and Line~\ref{lkcsr-line:group-update-1}.
If not, then we set $A_i$ to $1$ (Line~\ref{lkcsr-line:group-update}).
The next level for $i$, stored in $L_{r+1}[i]$ is then determined by $A_i$ (Line~\ref{lkcsr-line:update-list}).

After iterating through all of the $\ceil{4\log^2 n}$ rounds, we use the final levels of every vertex to estimate
its core number. For each vertex $i$, we use $L_{\ceil{\numlevels}-1}[i]$ which is the level that vertex $i$ is on
at the end of the procedure (Line~\ref{newcoren:max-g}). Then, we calculate the core number estimate using the
equation in Line~\ref{newcoren:estimate-equation}.
Using the improved analysis for the
MAT framework, we can obtain our desired error bounds.

\begin{algorithm}[htb!]
\caption{Improved $\eps$-LEDP $k$-Core Decomposition in $O(\log^2 n)$ Rounds Based on~\cite{DLRSSY22}}
\label{alg:improved-ledp-kcore}
\textbf{Input:} Graph $G=(V,E)$, privacy parameter $\eps>0$.\\
\textbf{Output:} $(2,c'\log(n)/\eps)$-approximate core numbers of each node in $G$ for sufficiently large constant $c' > 0$\\
\begin{algorithmic}[1]
\STATE Set $\psi = 0.1\coren$ and $\lambda = \frac{2(30- \eta)\eta}{(\eta + 10)^2}$.\\
\FOR{$i = 1$ to $n$}
    \STATE $\tilde{\ell}(i) \leftarrow  \text{Lap}(4/\eps)$.\label{lkcsr-line:noise-1}
    \STATE $L_r[i] \leftarrow 0$ for all $r \in \left[\ceil{4\log^2 n}\right]$.\label{lkcsr-line:initialize-L}
\ENDFOR
\FOR{$\lcur = 0$ to $\ceil{\numlevels} - 1$}\label{lkcsr-line:iterate}
\FOR{$i = 1$ to $n$}\label{lkcsr-line:iterate-vertex}
    \STATE $L_{\lcur + 1}[i] \leftarrow L_{\lcur}[i]$.\label{lkcsr-line:old-level}\\
    \STATE $A_i \leftarrow 0$.
    \IF{$L_\lcur[i] = \lcur$}\label{lkcsr-line:level-cur-level}
        \STATE Let $\nup_{i}$ be the number of neighbors $j \in \adj_i$ ($\adj_i$ is $i$'s adjacency list) where $L_\lcur[j] =
                \lcur$.\label{lkcsr-line:compute-up}\\
        \STATE Sample $X \leftarrow  \text{Lap}(8/\eps)$.\label{lkcsr-line:sample-noise}\\
        \STATE Compute $\hnup_{i} \leftarrow \nup_{i} + X$.\label{lkcsr-line:compute-noisy-up}\\
        \IF{$\hnup_{i} \leq \upexp^{\lfloor r/(\numgrouplevels)\rfloor} + \tilde{\ell}(i)$}\label{lkcsr-line:move-up-condition}
            \STATE Stop all future queries for $i$ and
            set $A_i \leftarrow 0$.\label{lkcsr-line:group-update-1}
        \ELSE
            \STATE Continue and set $A_i \leftarrow 1$.\label{lkcsr-line:group-update}\\
        \ENDIF
    \ENDIF
    \STATE $i$ \release $A_i$.\label{lkcsr-line:release-ai}\\
    \STATE $L_{\lcur + 1}[i] \leftarrow L_{\lcur}[i] + 1$ for every $i$ where $A_i = 1$.\label{lkcsr-line:update-list}
\ENDFOR
\STATE Curator publishes $L_{\lcur + 1}$.\label{lkcsr-line:levels}\\ %
\ENDFOR
\STATE $C \leftarrow \emptyset$.\label{lkcsr-line:c-empty}\\
\FOR{$i = 1$ to $n$}
    \STATE Let $\ell'\leftarrow L_{\ceil{\numlevels}-1}[i]$.\label{newcoren:max-g}\\
    \STATE $\kest(i)\leftarrow\upexpold^{\max\left(\lfloor\frac{\ell' + 1}{4\lceil\log_{1+\psi}(n)\rceil}\rfloor-1, 0\right)}$.\label{newcoren:estimate-equation}
    \STATE $C \leftarrow C \cup \{\kest(i)\}$.\label{newcoren:compute-estimate}
\ENDFOR
\STATE \textbf{Release} $C$.\label{lkcsr-line:return-noisy-density}
\end{algorithmic}
\end{algorithm}

We first show that
~\cref{alg:improved-ledp-kcore} is $\eps$-LEDP.

\begin{lemma}\label{lem:2-approx-ledp}
    \cref{alg:improved-ledp-kcore} is $\eps$-local edge differentially private.
\end{lemma}

\begin{proof}
    We formulate~\cref{alg:improved-ledp-kcore} in the context of~\cref{alg:local multidimensional Above Threshold} which implies
    that~\cref{alg:improved-ledp-kcore} is $\eps$-LEDP. As in the proof of~\cref{thm:optimal-privacy}, we use an instance of the
    MAT mechanism with input graph $G$, privacy parameter $\eps$, and threshold vector $\overrightarrow{T}$. We now adaptively define the
    queries.

    For each round $r$, the $r$-th threshold
    consists of $\upexp^{\lfloor r/(\numgrouplevels)\rfloor} - U_i$. The sensitivity of the
    vector containing such queries for each $v \in V$ is $2$ for each round $r$  (so $D = 2$).
    Let $q_r(i)$ be a variable %
    that denotes the answer to the adaptive
    query from round $r$ for node $i$. Then, we can compute $A_i$ (Lines~\ref{lkcsr-line:group-update-1} and~\ref{lkcsr-line:group-update})
    (via post-processing) for each node $i$ using only $q_r(i)$ and
    $L_{\lcur + 1}[i]$ is computed from $A_i$ and inductively from $L_{\lcur}[i]$. The approximate core numbers can be
    obtained from the $L_{\lcur}[i]$ values from each round
    via post-processing as shown in the original proof in~\cite{DLRSSY22}.
    Hence, our algorithm can be implemented using MAT and
    by~\cref{thm:mat-ledp} and~\Cref{lem:post} is $\eps$-LEDP.
\end{proof}

\begin{theorem}\label{thm:approx-logn-rounds}
    \cref{alg:improved-ledp-kcore} is $\eps$-LEDP and outputs $\left(2+\eta, O\left(\frac{\log n}{\eps}\right)\right)$-approximate core numbers $\hat{k}(v)$
    with probability at least $1 - O\left(\frac{1}{n^c}\right)$ for any constant $c \geq 1$ in $O(\log^2 n)$ rounds.
\end{theorem}

We defer the proof of~\cref{thm:approx-logn-rounds} to~\cref{app:distributed-approx-proof} since it is nearly
identical to the proof of the original algorithm.

Note that using the MAT technique in~\cref{alg:multidimensional Above Threshold}, we can also remove a factor of $O(\log n)$ in the additive error using the $O(\log n)$ algorithm given in Algorithm 3 of~\cite{DLRSSY22}; however, $\Delta = 2\log n$ for that algorithm when reframed in the MAT framework. Thus, we would obtain a $O(\log n)$ round algorithm which is $\eps$-LEDP but gives $O\left(2 + \eta, O\left(\frac{\log^2 n}{\eps}\right)\right)$-approximate core numbers.

\subsection{Lower bounds for core decomposition}
\label{sec:non-interactive}

Let $\eps > 0$ be any positive constant. We first give a lower bound in the centralized model (that carries over to the local model) and then give a lower bound for the local model.

\label{app:packinglb}

\begin{theorem}
\label{thm:coreness-centralized-lb}
Let $\gamma \ge 1$ be a constant and let $V$ be a set of $n$ vertices. Suppose that $\mathcal{M}$ is an $\eps$-edge differentially private mechanism in the centralized model that estimates the coreness of every vertex in a given graph $G$ on $V$ such that, for all vertices $v \in V$ with actual coreness $k(v)$, the estimated coreness $\tilde{k}(v)$ of $v$ satisfies
\[
\gamma^{-1} k(v) - \alpha \le \tilde{k}(v) \le \gamma k(v) + \alpha  \text { for all $v \in V$ simultaneously with probability at least $p$. }
\]
    Then $\alpha = \Omega(\gamma^{-1}\log (np) / \eps)$.
\end{theorem}
\begin{proof}
Let $d = \lceil (2\alpha+1)\gamma\rceil$  and let $G$ be any $(d+1)$-regular graph on $V$. For each vertex $v \in V$, let $G_v$ be the same as $G$, except that all $d+1$ edges incident to $v$ are removed. Observe that, in $G_v$, $v$ has coreness 0, while all other vertices $u \neq v$ have coreness at least $d$. Let $\mathcal{C}_v$ be the set of all coreness estimate vectors $\tilde{k}$ such that $\tilde{k}(v) \leq \alpha$ and $\tilde{k}(u) > \alpha$ for all $u \neq v$. Observe that $\mathcal{C}_v$ and $\mathcal{C}_u$ are disjoint for $u \neq v$. Furthermore, since $\gamma^{-1}d - \alpha = \alpha+1$,  $\mathcal{M}(G_v)$
must return $\tilde{k}(v) \le \alpha$ and for all $u \ne v$, $\tilde{k}(u) \ge \alpha+1$ , i.e., an estimate in $\mathcal{C}_v$ with probability at least $p$.

Since $G$ and $G_v$ differ by $d+1$ edges and $\mathcal{M}$ satisfies $\eps$-edge differential privacy, we have that $\Pr( \mathcal{M}(G) \in \mathcal{C}_v ) \geq e^{-\eps (d+1)} \Pr( \mathcal{M}(G_v) \in \mathcal{C}_v ) \geq e^{-\eps (d+1)} p$. Since $\mathcal{C}_u$ and $\mathcal{C}_v$ are disjoint for $u \neq v$,
$1 \geq \Pr( \bigcup_{v \in V} (\mathcal{M}(G) \in \mathcal{C}_v) ) = \sum_{v \in V} \Pr( \mathcal{M}(G) \in \mathcal{C}_v ) \geq  n e^{-\eps (d+1)} p$.
By rearranging, $\lceil (2\alpha+1)\gamma \rceil = d \geq \ln (np)/\eps-1$. Therefore, $\alpha = \Omega(\gamma^{-1}\log (np)/\eps)$.
\end{proof}

Let $V$ be a set of $2n+1$ vertices and let $x \in V$ be a fixed vertex. Consider any $\eps$-edge differentially private mechanism $\mathcal{M}$ that \emph{non-interactively} (i.e., in a single round) estimates the coreness of $x$ in a given graph on $V$, in the local model. We show that there is a large family of graphs on which $\mathcal{M}$ has constant probability of returning an estimate with $\Omega(\sqrt{n})$ error for $x$.

\begin{theorem}
\label{thm:non-interactive-coreness-lb}
For any constant $\eps > 0$, there exists a constant $0 < \eta < \frac{1}{2}$ such that the following holds. Suppose that $\mathcal{M}$ is a non-interactive $\eps$-edge differentially private local mechanism that estimates the coreness of a fixed vertex, $x$, in an arbitrary $n$-vertex graph such that, if $k(x)$ is the actual coreness of $x$, then the estimated coreness $\tilde{k}(x)$ of $x$ satisfies:
\[
    k(x) - \alpha \leq \tilde{k}(x) \leq k(x) + \alpha \text{ with  probability at least $\tfrac{1}{2}+\eta$. }
\]
Then there is a family of $n$-vertex graphs of size $2^{\Omega(n)}$ on which $\alpha = \Omega(\sqrt{n})$.
\end{theorem}

Our approach is to reduce to a known lower bound on the error of privately answering a linear number of random \emph{inner product queries} on a secret dataset $X \in \{0,1\}^n$. Here, two datasets $X$ and $X'$ are \emph{neighboring} if they differ in at most one coordinate, a query is specified by a vector $Q\in \mathbb{R}^n$, and the \emph{error} of a response $r$ to query $Q$ is $|r - \langle Q,X\rangle|$. \cite{ELRS22} were the first to use this approach to prove lower bounds in the local model.

Roughly speaking, the lower bound says that no  $\eps$-differentially private mechanism (with a trusted curator) can answer $O(n)$ random inner product queries in $\{-1,1\}^n$ so that, with constant probability, a large fraction of the answers have $o(\sqrt{n})$ error. (Otherwise, an ``attacker'' can use such queries to ``reconstruct'' $X$ with high accuracy, violating privacy.) We formally state the lower bound in the next subsection and use it to prove the following modified variant, adapted to inner product queries in $\{0,1\}^n$. The idea is to convert a mechanism answering inner product queries in $\{0,1\}^n$ into a mechanism answering inner product queries in $\{-1,1\}^n$ with roughly the same error and privacy loss.

\begin{restatable}{theorem}{reconstruction}
\label{thm:reconstruction-attack}
    For any constants $\eps > 0$ and $\frac{1}{20} > \delta > 0$, there is a constant $0 < \eta < \frac{1}{2}$ such that no $(\eps,\delta)$-differentially private mechanism can answer $m = O(n)$ random inner product queries in $\{0,1\}^n$ on a secret dataset $X \in \{0,1\}^n$ such that, with probability at least $\Omega(\sqrt{\delta})$, a $(\frac{1}{2} + \eta)$-fraction of its answers have $o(\sqrt{n})$ error.
\end{restatable}

\subsection{Inner product queries}

\cite{de2012lower} proved a lower bound on the additive error of any differentially private mechanism that answers $m = O(n)$ random inner product queries in $\{-1,1\}^n$ on a secret dataset in $\{0,1\}^n$. Specifically, the lower bound says that if the error is $O(\sqrt{n})$ on a $(\frac{1}{2}+\eta)$-fraction of the responses with probability $\Omega(\sqrt{\delta})$, then the mechanism is not $(\eps,\delta)$-differentially private.

\begin{theorem} [Theorem 4.1 of \cite{de2012lower}]
    For any $n \in \mathbb{N}$, $\eps > 0$, and $1/20 > \delta > 0$, there exists positive constants $\alpha$, $\gamma$, and $\eta < 1/2$ such that any mechanism $\mathcal{M}$ that answers $m = \alpha n$ random inner product queries $Q^{(1)},\dots,Q^{(m)} \in \{-1,1\}^n$ on a secret dataset $X\in \{0,1\}^n$ satisfying
    \[
        \Pr_{\mathcal{M}, Q^{(1)},\dots,Q^{(m)}} \left[ \Pr_{i \in [m]} [ |\mathcal{M}(X)_i - \langle X, Q^{(i)} \rangle| \leq \gamma \sqrt{n} ] \geq \frac{1}{2} + \eta \right] \geq 3\sqrt{\delta} \,
    \]
    where $\mathcal{M}(X)_i$ denotes the response of $\mathcal{M}$ on query $Q^{(i)}$, is not $(\eps,\delta)$-differentially private.
\end{theorem}

To obtain the lower bound on inner product queries in $\{0,1\}^n$ that we desire, we show that any $\eps$-differentially private mechanism $\tilde{\mathcal{M}}$ that answers $m$ random inner product queries in $\{0,1\}^n$ can be converted into a $2\eps$-differentially private mechanism $\mathcal{M}$ that answers $m$ random inner product queries in $\{-1,1\}^n$. The idea is to observe that, for any $Q \in \{-1,1\}^n$, we may write $\langle Q,X \rangle = 2\langle \tilde{Q}, X \rangle - \langle \mathbf{1}, X \rangle$, where $\tilde{Q} \in \{0,1\}^n$ is such that $\tilde{Q}_i = \frac{1}{2}(\tilde{Q}_i + 1)$ and $\mathbf{1} \in \{1\}^n$ is the all-ones vector. Using this, we can simulate $\tilde{\mathcal{M}}$ on $X$, scale each response by a factor of 2, and then subtract a noisy version of $\langle \mathbf{1},X\rangle$ from each response. More precisely, to answer $m$ random inner product queries in $\{-1,1\}^n$ on $X \in \{0,1\}^n$, $\mathcal{M}$ does the following:
\begin{enumerate}
    \item Generate $y \sim \mathsf{Lap}(1/\eps)$ and release $\tilde{x} = \langle \mathbf{1}, X \rangle + y$.
    \item Run $\mathcal{M}$ on $X$ to obtain responses $(r_1,r_2,\dots,r_m)$.
    \item Return $(2r_1 - \tilde{x}, 2r_2 - \tilde{x}, \dots, 2r_m - \tilde{x})$.
\end{enumerate}

The first step is $\eps$-differentially private by Lemma~\ref{lem:Laplacemech}. The second step is $\eps$-differentially private by assumption. Therefore, by composition, the first two steps are $2\eps$-differentially private. The last step is simply post-processing.

Notice that since $\mathcal{M}$ answers $m$ random inner product queries $\tilde{Q}^{(1)},\dots,\tilde{Q}^{(m)} \in \{0,1\}^n$ on $X$, the vectors $Q^{(1)},\dots,Q^{(m)} \in \{-1,1\}^n$, where $Q_i^{(j)} = 2\tilde{Q}_i^{(j)} + 1$ are random. Furthermore, the error of the $j$th answer is $|(2r_j - \tilde{x}) - \langle Q^{(j)}, X\rangle| \leq |2r_j - 2\langle \tilde{Q}^{(j)}, X \rangle| + |2\langle \tilde{Q}^{(j)}, X \rangle - \langle \mathbf{1}, X \rangle - \langle Q^{(j)}, X \rangle| + |y| = 2|r_j - \langle \tilde{Q}^{(j)}, X \rangle| + |y|$. By Laplace tail bounds (Lemma~\ref{lem:laplace_tailbounds}), $|y| \leq O(\eps^{-1}\log n)$ with high probability. Therefore, if the error of $j$th answer of $\tilde{\mathcal{M}}$ is $o(\sqrt{n})$, then so is the error of the $j$th answer of $\mathcal{M}$, assuming $\eps$ is a constant. This gives us \cref{thm:reconstruction-attack}.

\subsection{Single round lower bound}

To apply Theorem~\ref{thm:reconstruction-attack}, we construct a $2\eps$-differentially private mechanism $\mathcal{N}$ that answers $m$ random inner product queries in $\{0,1\}^n$. For each query, $\mathcal{N}$ simulates $\mathcal{M}$ on a \emph{query graph} in which the coreness of $x$ is roughly the intended answer to the query. We show that, with constant probability, a large fraction of $\mathcal{N}$'s responses will have the same error as $\mathcal{M}$. Therefore, $\mathcal{M}$ has $\Omega(\sqrt{n})$ error with constant probability. 

\subparagraph*{Query graphs} Fix a partition $(A,B)$ of $V \setminus \{x\}$ with $|A| = |B|$ and an enumeration, $a_1,\dots,a_{n}$, of the vertices in $A$. Let $Q \in \{0,1\}^n$ be an arbitrary inner product query.  The \emph{query graph} for $Q$ on dataset $X$ is the graph $G_X(Q)$ on $V$ defined as follows.
\begin{enumerate}
    \item For all $i \in [n]$, $X_i \in \{0,1\}$ indicates whether $a_i$ is adjacent to $x$ and $Q_i \in \{0,1\}$ indicates whether $a_i$ is adjacent to (0) no vertex in $B$ or (1) every vertex in $B$.
    \item The vertices of $B$ form a clique, which $x$ is not adjacent to.
\end{enumerate}
The idea is that $Q$ is used to bound the coreness of vertices in $A$. Specfically, if $Q_i = 1$, then $a_i$ is adjacent to every vertex in $B$ and, hence, has coreness at least $n-1$. Otherwise, $a_i$ can only be adjacent to $x$, so it has coreness at most $1$. Thus, the coreness of $x$ is roughly the number of neighbors $a_i \in A$ such that $Q_i = 1$, i.e., $\langle Q, X \rangle$. (If $Q$ is the all-zero vector, then the coreness of $x$ may still be 1.)

\begin{restatable}{lemma}{corenessquerygraph}
\label{lem:coreness-query-graph}
    The coreness of $x$ in $G_X(Q)$ is either $\langle Q, X \rangle$ or $\langle Q,X \rangle +1$.
\end{restatable}
\begin{proof}
Let $A' = \{ a_i \in A \mid Q_i = X_i = 1 \}$. Notice that $|A'| = \langle Q, X \rangle$. If $|A'| = 0$, then every neighbor (if any) of $x$ has degree 1. Hence, the coreness of $x$ is at most $1 = \langle Q,X \rangle +1$.

Now suppose $|A'| \neq 0$. We will show that the coreness of $x$ equals $|A'| = \langle Q,X \rangle$. By construction, every vertex in $A'$ is adjacent to $x$ and every vertex in $B$. It follows that $A' \cup B \cup \{x\}$ induces a subgraph of $G_X(Q)$ in which every vertex has (induced) degree at least $|A'| \geq 1$. Hence, the coreness of $x$ is at least $|A'|$.

Next consider any set of vertices $S$ containing $x$. If $S$ contains a vertex $a_i \in A \setminus A'$ that is adjacent to $x$, then $Q_i = 0$ and the degree of $a_i$ in the subgraph of $G_X(Q)$ induced by $S$ is $1 \leq |A'|$. Otherwise, every neighbor of $x$ in the subgraph of $G_X(Q)$ induced by $S$ is in $A'$ and, hence, the induced degree of $x$ is at most $|A'|$. Hence, in either case the coreness of $x$ is at most $|A'|$.
\end{proof}

\subparagraph*{Answering random queries}
Na\"{i}vely answering queries by simulating $\mathcal{M}$ on each of the corresponding query graphs leads to prohibitively large privacy loss. Our key observation is that the vertices whose neighborhoods in the query graphs depend on the secret dataset $X$, namely $\{x\} \cup A$, only have a few possible neighborhoods among all query graphs on $X$. This suggests that, to reduce the privacy loss, we can generate only a few messages for these vertices and \emph{reuse} them in all simulations.

\begin{observation}
\label{obs:querygraphs}
    The following holds for any secret dataset $X \in \{0,1\}^n$.
\begin{enumerate}
    \item The neighbors of $x$ are the same in all query graphs on $X$.
    \item Each vertex in $A$ only has two possible neighborhoods among all query graphs on $X$ (namely, $\{x\}$ and $B \cup \{x\}$, if $X_i = 1$, and $\emptyset$ and $B$ otherwise).
    \item For all datasets $X$, the neighborhood of each vertex $b \in B$ in the query graph $G_X(Q)$ is a function of $Q$, i.e., $(B \setminus \{b\}) \cup \{ a_i \in A \mid Q_i = 1 \}$.
\end{enumerate}
\end{observation}

Specifically, our mechanism $\mathcal{N}$ answers $m$ random inner product queries $Q^{(1)},\dots,Q^{(m)} \in \{0,1\}^n$ on a secret dataset $X \in \{0,1\}^n$ as follows.
\begin{enumerate}
    \item Let $\mathbf{0} \in \{0,1\}^n$ and $\mathbf{1} \in \{0,1\}^n$ be the all-zeros and all-ones vectors, respectively.
    \begin{enumerate}
        \item Run the user algorithm of $x$ (specified by $\mathcal{M}$) on its neighborhood in $G_X(\mathbf{0})$ to obtain the message $\pi_x$.
        \item For each $i \in [n]$: run the user algorithm of $a_i$ (specified by $\mathcal{M}$) on its neighborhood in $G_X(\mathbf{0})$ and $G_X(\mathbf{1})$ to obtain the messages $\pi_{a_i}(0)$ and $\pi_{a_i}(1)$, respectively.
    \end{enumerate}
    \item For each query $Q^{(j)}$, \emph{simulate} a run of $\mathcal{M}$ on $G_X(Q^{(j)})$ as follows:
    \begin{enumerate}
        \item Run the user algorithm of each vertex $b \in B$ (specified by $\mathcal{M}$) on its neighborhood in the query graph $G_X(Q^{(j)})$ to obtain the message $\pi_b^{(j)}$.
        \item Run the server's algorithm (specified by $\mathcal{M}$) on the transcript \[\{\pi_x\} \cup \{\pi_{a_i}(Q^{(j)}_i) \mid i \in [n] \} \cup \{\pi^{(j)}_b \mid b \in B \} \] to obtain an estimate $\tilde{k}_j(x)$ of the coreness of vertex $x$ in the query graph $G_X(Q^{(j)})$.
    \end{enumerate}

    \item Return $(\tilde{k}_j(x) : j \in [m])$.
\end{enumerate}

\subparagraph*{Analysis}

Since we run the user algorithm of each vertex in $\{x\} \cup A$ on its respective neighborhood in at most two graphs and $\mathcal{M}$ is $\eps$-edge differentially private, the messages generated in the first step are collectively $2\eps$-edge differentially private. Since the pairs of graphs $G_X(\mathbf{0}), G_{X'}(\mathbf{0})$ and $G_X(\mathbf{1}), G_{X'}(\mathbf{1})$ each differ in at most one edge when datasets $X$ and $X'$ are neighboring, it follows that the first step is $2\eps$-differentially private. The other steps can be viewed as simply post-processing the outputs (messages) of the first step. Therefore, the entire procedure is $2\eps$-differentially private.

\begin{restatable}{lemma}{lbdp}
    $\mathcal{N}$ is $2\eps$-differentially private.
\end{restatable}
\begin{proof}
Let $X, X' \in \{0,1\}^n$ be two datasets that differ only at the $i$th coordinate.    Consider any valid combination of messages, $\{\pi_x\} \cup \{\pi_{a_j}(0) \mid j \in [n] \} \cup \{ \pi_{a_j}(1) \mid j \in [n] \}$, generated in the first step. Since $X$ and $X'$ differ only at the $i$th coordinate, the neighborhoods of $x$ in $G_X(\mathbf{0}),G_{X'}(\mathbf{0})$ differ in only one edge, namely $\{x,a_i\}$.
Similarly, for the neighborhoods of $a_i$ in the pairs of graphs $G_X(\mathbf{0}),G_{X'}(\mathbf{0})$ and $G_X(\mathbf{1}),G_{X'}(\mathbf{1})$. The neighborhoods of the other vertices in $A$ are the same in each pair of graphs $G_X(\mathbf{0}), G_{X'}(\mathbf{0})$ and $G_{X}(\mathbf{1}),G_{X'}(\mathbf{1})$. It follows that the ratio of the probabilities of seeing the (partial) transcript $\{\pi_x\} \cup \{ \pi_{a_j}(0) \mid j \in [n] \}$ on the graphs $G_{X}(\mathbf{0}),G_{X'}(\mathbf{0})$ is at most $e^{\eps}$, since the transcript is $\eps$-edge differentially private.
Similarly, for the ratio of the probabilities of seeing the (partial) transcript $\{ \pi_{a_j}(1) \mid j \in [n] \}$ on the graphs $G_X(\mathbf{1}), G_{X'}(\mathbf{1})$. Thus, the ratio of the probabilities of seeing the entire combination of messages in the first step on datasets $X$ and $X'$ is at most $e^{\eps} \cdot e^{\eps} = e^{2\eps}$ and the first step is $2\eps$-differentially private.
The remaining steps do not require knowledge of $X$ and, hence, only serve as post-processing. Therefore, $\mathcal{N}$ is $2\eps$-differentially private.
\end{proof}

Lemma~\ref{lem:coreness-query-graph} implies that $\mathcal{N}$ produces a response to a query with additive error exceeding $\alpha+1$ only if the simulation of $\mathcal{M}$ on the corresponding query graph produces an estimate of the coreness of $x$ that has additive error exceeding $\alpha$. We note that, since the messages of vertices in $\{x\} \cup A$ are reused, the simulations are \emph{not} independent. Nonetheless, Markov's inequality can be used to bound the probability that a large number of simulations have error exceeding $\alpha$.

\begin{restatable}{lemma}{lbacc}
    Suppose that $\mathcal{M}$ has error exceeding $\alpha$ with probability at most $\beta$. Then, with probability at least $1-1/\gamma$, at least a $(1-\gamma\beta)$-fraction of the responses of $\mathcal{N}$ have error at most $\alpha+1$.
\end{restatable}
\begin{proof}
    For $j \in \{1,\dots,m\}$, let $X_j \in \{0,1\}$ be the indicator random variable for whether the response to the $j$th query has error exceeding $\alpha+1$, i.e., $|\tilde{k}_j(x) - \langle Q^{(j)}, X \rangle|>\alpha+1$. By Lemma~\ref{lem:coreness-query-graph}, if $k_j(x)$ is the actual coreness of $x$ in $G_X(Q^{(j)})$, then $|k_j(x) - \langle Q^{(j)},X \rangle| \leq 1$. Hence, by triangle inequality, $|\tilde{k}_j(x) - \langle Q^{(j)}, X \rangle| \leq |\tilde{k}_j(x) - k_j(x)| + |k_j(x) - \langle Q^{(j)}, X \rangle| \leq |\tilde{k}_j(x) - k_j(x)| + 1$. Since $\mathcal{M}$ has error exceeding $\alpha$ with probability at most $\beta$, it follows that $E[X_j] \leq \beta$. Hence, by linearity of expectation, $E[\sum_{j=1}^m X_j] = \sum_{j=1}^m E[X_j] \leq \beta m$ and, by Markov's inequality, $\Pr(\sum_{i=1}^m X_i \geq \gamma\beta m) \leq 1/\gamma$.
    Therefore, with probability $1-1/\gamma$, at least $(1 - \gamma\beta) m$ responses of $\mathcal{N}$ have error at most $\alpha+1$.
\end{proof}

To summarize, we have shown that, if $\mathcal{M}$ estimates the coreness of a fixed vertex with error exceeding $\alpha$ with probability at most $\beta = \frac{1}{2}-\eta$, then with constant probability (say $1-1/1.001 \approx 0.001$), a large fraction ($1-1.001\beta \approx \frac{1}{2}+\eta$) of the responses of $\mathcal{M}$ have error at most $\alpha + 1$. Since $\mathcal{N}$ is $2\eps$-differentially private, it is $(2\eps,\delta)$-differentially private for any $\delta > 0$. Therefore, by the contrapositive of Theorem~\ref{thm:reconstruction-attack}, $\alpha = \Omega(\sqrt{n})$.
\section{LEDP Densest Subgraph}

\subsection{Algorithms from MAT}

We present our $(2, O(\log(n)/\eps))$-approximate
and $(4+\eta, O(\log(n)/\eps))$-approximate
edge-differentially private densest subgraph algorithms
in this subsection.
Our algorithms follow from our $k$-core decomposition algorithm and the well-known folklore theorem that the density of the
densest subgraph, $\density^*$, falls between half of the maximum $k$-core value and the maximum $k$-core value. In fact, one can show
that the $(k_{\max}/2)$-core (where $k_{\max}$ is the maximum value of a non-empty core)
contains the densest subgraph in $G$ (see e.g.\ Corollary 2.2 of~\cite{SLDS23}).
Furthermore, the $k_{\max}$-core gives a $2$-approximation of the densest subgraph in the non-private setting.

We first present our $(2, O(\log(n)/\eps))$-approximation algorithm in~\cref{alg:densest subgraph algorithm}.
We use~\cref{alg:optimal-algorithm} to obtain differentially private approximations of the $k$-core values of each vertex.
Then, we find the maximum returned core value, denoted by $\hat{k}_{\max}$. Finally, we take the induced subgraph
consisting of all vertices with approximate core values at least $\hat{k}_{\max} - \frac{c'\log(n)}{\eps}$ for sufficiently
large constant $c'$. This means that
we include all $k_{\max}$ vertices with high probability, and potentially some vertices with core values at least $k_{\max} -
\frac{2c'\log(n)}{\eps}$. However, we prove that this still results in a good approximate densest subgraph below.

Similarly, using our low round $(2+\eta, O(\log n/\eps))$-approximate $k$-core algorithm instead gives us a $(4+\eta, O(\log n/\eps)$-approximate densest subgraph algorithm, which is presented in \cref{alg:lowround densest subgraph algorithm}.

\begin{algorithm}[h]
\caption{$\eps$-LEDP $(2, O(\log n/\eps)$-approximate Densest Subgraph in $n$ rounds}
\label{alg:densest subgraph algorithm}
\textbf{Input:} Graph $G=(V,E)$, privacy parameter $\eps>0$.\\
\textbf{Output:} A $(2,O(\log(n)/\eps))$-approximate densest subgraph $S\subseteq V$\\
\begin{algorithmic}[1]
\STATE Run \Cref{alg:optimal-algorithm} with parameter $\eps$ to obtain approximate core numbers $\hat{k}(v)$ for each $v\in V$
\STATE Let $c'$ be such that $\hat k(v)$ satisfy $k(v) - c' \log n /\eps \leq \hat{k}(v) \leq k(v) + c' \log n / \eps$ with probability $\ge 1 - n^{-c}$ for a fixed constant $c > 1$.
\STATE Let $\hat{k}_{\max} \leftarrow \max_{v \in V}\left(\hat{k}(v)\right)$
\STATE Find $S \leftarrow \{v \mid \hat{k}(v) \geq \hat{k}_{\max} - c' \log n /\eps\}$
\STATE Return $S$
\end{algorithmic}
\end{algorithm}

\begin{algorithm}[h]
\caption{$\eps$-LEDP $(4+\eta, O(\log n/\eps)$-approximate Densest Subgraph in $O(\log^2 n)$ rounds}
\label{alg:lowround densest subgraph algorithm}
\textbf{Input:} Graph $G=(V,E)$, privacy parameter $\eps>0$.\\
\textbf{Output:} A $(4+\eta,O(\log(n)/\eps))$-approximate densest subgraph $S\subseteq V$\\
\begin{algorithmic}[1]
\STATE Run \Cref{alg:improved-ledp-kcore} with parameter $\eps$ to obtain approximate core numbers $\hat{k}(v)$ for each $v\in V$
\STATE Let $c'$ be such that $\hat k(v)$ satisfy $k(v) - c' \log n /\eps \leq \hat{k}(v) \leq (2+\eta) \cdot k(v) + c' \log n / \eps$ with probability $\ge 1 - n^{-c}$ for a fixed constant $c > 1$.
\STATE Let $\hat{k}_{\max} \leftarrow \max_{v \in V}\left(\hat{k}(v)\right)$
\STATE Find $S \leftarrow \{v \mid \hat{k}(v) \geq \hat{k}_{\max} / (2 + \eta) - c' \log n /\eps\}$
\STATE Return $S$
\end{algorithmic}
\end{algorithm}

Since selecting the subset $S$ in both algorithms is simply post-processing of the approximate $k$-core values, the entire procedure has the same differential privacy guarantees as the coreness estimation mechanism. The next lemma gives sufficient conditions for the accuracy of the estimation to scale with the accuracy of the underlying coreness estimation procedure. While~\cite{DLRSSY22} implicitly proves this, we show this explicitly below.

\begin{lemma}
\label{lem:density-coreness-accuracy}
    Let $\gamma \geq 1$ be a constant. Suppose that for each vertex $v \in V$ with actual coreness $k(v)$:
    \begin{itemize}
        \item the estimated coreness $\hat{k}(v)$ of $v$ satisfies $k(v) - \alpha \leq \hat{k}(v) \leq \gamma \cdot k(v) + \alpha$ and
        \item every vertex in $G[U]$  has (induced) degree at least $\hat{k}(v)/\gamma - \alpha$, where $U$ is the set of all vertices $u$ with estimated coreness $\hat{k}(u) \geq \hat{k}(v)$.
    \end{itemize}
    Then the density of $G[S]$ is at least $\density^*/2\gamma - O(\alpha)$ where $S = \{ v \mid \hat k(v) \ge \hat k_{\max} / \gamma - \alpha \}$.
\end{lemma}
\begin{proof}
By the first assumption, 
\[
\hat{k}_{\max} = \max_{v \in V} \hat{k}(v) \geq \max_{v \in V} k(v) - \alpha = k_{\max} - \alpha
\]
By the second assumption, every vertex $v$ in $G[S]$ has induced degree at least 
\[
d_S(v) \ge 
\frac{\hat{k}_{\max}}{\gamma} - \alpha 
\geq \frac{k_{\max}}{\gamma} - \left( \frac{1}{\gamma} + 1 \right)\alpha
\]
Counting the total degree of vertices inside $G[S]$ and using the handshake lemma, we get
\[
|E(S)|
\ge |S| \cdot \left( \frac{k_{\max}}{\gamma} -  \left( \frac{1}{\gamma} + 1 \right)\alpha \right) \cdot \frac{1}{2}
\]
Dividing by $|S|$ gives a lower bound on the density of $G[S]$.
Since $\rho^* \le k_{\max}$, this is at least $\density^*/2\gamma - O(\alpha)$.
\end{proof}

Since we use our coreness estimation mechanisms from Algorithms~\ref{alg:optimal-algorithm} and~\ref{alg:improved-ledp-kcore}, which both satisfy the conditions of Lemma~\ref{lem:density-coreness-accuracy} (see Theorems~\ref{thm:optimal-utility} and~\ref{thm:approx-logn-rounds}), we immediately obtain the following.

\begin{restatable}{theorem}{densestsubgraph}
\label{thm:densestsubgraph}
For any $\eps, \eta > 0$, there are local $\eps$-edge differentially private mechanisms (\cref{alg:densest subgraph algorithm} and \cref{alg:lowround densest subgraph algorithm}) that return subsets of vertices $S$ and $S'$ from a given $n$-vertex graph $G$, respectively, such that:
\begin{itemize}
    \item \cref{alg:densest subgraph algorithm} and \cref{alg:lowround densest subgraph algorithm} run for at most $n$ and $O(\log^2 n)$ rounds, respectively, in any execution.
    \item With probability $1-n^{-c}$ for a fixed constant $c > 1$, the densities of $G[S]$ and $G[S']$ is at least $\density^*/2 - O(\log n / \eps)$ and $\density^*/(4+\eta) - O(\log n / \eps)$, respectively, where $\density^*$ is the optimal density.
\end{itemize}
\end{restatable}

We also obtain an algorithm in the non-interactive model, which we state below.

\subsection{Single Round Algorithm}

We give a single round algorithm that estimates the densest subgraph upto an $O(\sqrt{n})$ additive error in the non-interactive model. It obtains the edges in the graph privately using Randomized Response and builds a \emph{randomized response graph} $G_R$. For each subset of vertices $S \subseteq V$, we estimate the number of edges inside $S$ as the unbiased estimator obtained from $G_R$. We show using a union bound over \emph{all} subsets that the estimated densities are at most an additive $O(\sqrt{n})$ factor away from the true densities. The key factor here is the normalization (by the size of the subset) in the definition of density, which reduces the variance from $|S|^2$ to constant.

\begin{algorithm}[ht]
\caption{Non-interactive single round private densest subgraph algorithm}
\label{alg:onerounddsg}
\textbf{Input:} Graph $G=(V,E)$, privacy parameter $\eps > 0$.\\
\textbf{Output:} A subset of vertices $S \subseteq V$ with large density.
\begin{algorithmic}[1]
\STATE Let $\text{id} = (1, \ 2, \ \ldots, \ n)$ be the identity permutation on the vertices, and set $p = e^{\eps}/(1+e^{\eps})$.
\STATE From each vertex $v$, receive an adjacency list $A_v$ of neighbouring edges to vertices \emph{after $v$ in permutation $\mathrm{id}$}, using Randomized Response with parameter $\eps$ for each edge.
\STATE Build a randomized response graph $G_{R}$ using the replies from each vertex.
\STATE Set $\Sout \gets \emptyset$, $\apx \density(\Sout) \gets 0$\;
\FOR{each $S \subseteq V$}
\STATE Let $E_R(S)$ be the number of edges inside $S$ in $G_R$.
\STATE Estimate the number of edges inside $S$ as $\apx E(S) = (2p-1)^{-1} \cdot \left( E_R(S) - \binom{|S|}{2} \cdot (1-p) \right)$
\STATE Estimate the density of $S$ as $\apx \density(S) = \apx E(S) / |S|$
\IF{$\apx \density(S) > \apx \density(\Sout)$}
\STATE Update $\Sout \gets S$
\ENDIF
\ENDFOR
\end{algorithmic}
\end{algorithm}

Recall that in Randomized Response, the true answer is returned with probability $p = e^{\eps}/1+e^{\eps}$, and the false answer with probability $1-p$.
The claimed running time in \cref{tab:results} and \cref{thm:one-round-ds} follows from the theorem below, since $\frac{e^x + 1}{e^x - 1} \le 1 + \frac{2}{x}$ for all $x > 0$.

\newcommand{\epart}{\ensuremath{\cdot (e^{\eps}+1)/(e^{\eps}-1)}}

\begin{restatable}{theorem}{onerounddsg}
\label{thm:onerounddsg}
Algorithm~\ref{alg:onerounddsg} is $\eps$-LEDP and, for any constant $c > 1$ with probability at least $1 - n^{-c}$, returns a subset of vertices $\Sout \subseteq V$ such that
\[
\density(\Sout) \ge \density^* - O(\sqrt{n} \epart )
\]
where $\density^*$ is the optimal density.
\end{restatable}

\begin{proof}
Note that for each pair $(u, v) \in V^2$, there is a single output from Randomized Response, which in particular is from the vertex that is earlier in the permutation id.
Thus, on two neighbouring graphs which differ in edge $(u,v)$, we only need to check the probability that the same answer is returned by Randomized Response on this edge on both graphs.
Since Randomized Response is $\eps$-DP (see for example \cite[Section 3.2]{dwork2014algorithmic}), Algorithm~\ref{alg:onerounddsg} is also $\eps$-LEDP.

Let $E(S)$ be the number of edges inside $S$ in $G$.
Our key goal will be to show that for each subset $S \subseteq V$,
\[
\abs{\apx E(S) - E(S)} \le O_{\eps}(|S| \cdot \sqrt{n}) \quad \text{ with probability $ \ge 1 -\frac{1}{n^c 2^n}$.}
\]
Then the lemma follows, since
\begin{align*}
&\Pr\left[\abs{ \density(\Sout) - \density^* } > 2 \sqrt{\left( n + c \ln n\right)} \epart  \right]\\
&\le \Pr\left[\exists \ S \subseteq V \text{ such that } \abs{  \apx \density(S) - \density(S)  } > 2 \sqrt{\left( n + c \ln n\right)} \epart \right] \numberthis \label{eq:existsbadsubset} \\
&\le \sum_{S \subseteq V} \Pr \left[ \abs{  \apx \density(S) - \density(S)  } > 2 \sqrt{\left( n + c \ln n\right)} \epart  \right] \\
&= \sum_{S \subseteq V} \Pr \left[ \abs{  \apx E(S) - E(S)  } > 2 \cdot |S| \cdot \sqrt{\left( n + c \ln n\right)} \epart \right] \\
&\le \sum_{S \subseteq V} \frac{1}{n^c \cdot 2^n} \numberthis \label{eq:edgeconcentration}
\le \frac{1}{n^c}
\end{align*}
where \cref{eq:existsbadsubset} follows, since for any $\kappa > 0$,
\[
\forall S \subseteq V, \abs{\apx \density(S) - \density(S)} \le \kappa \implies \abs{\max_{S \subseteq V} \apx \density(S) - \max_{S \subseteq V} \density(S)} \le \kappa
\]
and we prove \cref{eq:edgeconcentration} below, setting $\alpha = |S| \cdot \sqrt{\left( n + c \ln n\right)} \epart$. 

We study the properties of the random graph $G_R$ first. Each edge in $G$ is present with probability $p = e^{\eps}/(1+e^{\eps})$, and each edge not in $G$ is present with probability $1-p$.
We study this graph by constructing two random graphs whose union is $G_R$. The first graph, $H$, samples each edge of $G$ independently with probability $p$. The second graph, $H^C$, samples each edge not in $G$ with probability $1-p$. Then, since the edge sets of $H$ and $H^C$ are disjoint, it follows that $G_R = H \uplus H^C$.

Fix a subset $S \subseteq V$.
We first show that $\apx E(S)$ is an unbiased estimator of $E(S)$, by calculating the expected number of edges inside $S$ in $G_R$.
\begin{align*}
\E[E_R(S)]
&= p \cdot E(S) + (1-p) \cdot \left(\binom{|S|}{2} - E(S)\right) \\
&= (2p-1) \cdot E(S) + (1-p) \cdot \binom{|S|}{2}
\end{align*}
Thus $\E[\apx E(S)] = E(S)$ whenever $p \ne 1/2$, which is equivalent to $\eps \ne 0$. 
Now that we have an unbiased estimator, we want to show that the estimator is $O_{\eps}(|S| \cdot \sqrt{n})$-close to its mean with very high probability.
Note that $E_R(S) = E_H(S) + E_{H^C}(S)$.
\begin{align*}
\Pr \left[ \abs{  \apx E(S) - E(S)  } > 2 \alpha \right]
&= \Pr \left[ \abs{ \frac{E_R(S) - (1-p) \cdot \binom{|S|}{2}}{2p-1}  - E(S) } > 2\alpha \right] \\
&= \Pr \left[ \abs{ E_R(S) - (1-p) \cdot \binom{|S|}{2} - (2p-1) \cdot E(S) } > 2 \cdot (2p-1)) \cdot \alpha \right] \\
&= \Pr \left[ \abs{ E_{H}(S) + E_{H^C}(S) - (1-p) \cdot \binom{|S|}{2} - (2p-1) \cdot E(S) } > 2 \cdot (2p-1) \cdot \alpha \right] \\
&\le \Pr \left[ \abs{ E_{H}(S) - p \cdot {E(S)} } > (2p-1) \cdot \alpha \right] \\
&\ \ +
\Pr \left[ \abs{  E_{H^C}(S) - (1-p) \cdot \left( \binom{|S|}{2} - E(S) \right)  } > (2p-1) \cdot \alpha \right]
\end{align*}
where the last inequality follows using the following inequality and a union bound: for any $x, y \in \mathbb R$ and $z \ge 0$, it holds that
\[
|x| \le z \text{ and } |y| \le z \implies |x+y| \le 2z 
\]
To bound the first term in the final expression we use a concentration bound for $H$, and to bound the second term we use a concentration bound for $H^C$:
A Chernoff bound shows that for a sequence of random variables $X_1, \ldots, X_k$ which take values in $\{ 0,1 \}$, if $X = \sum_i X_i$ and $\mu = \E[X]$, then
\[
\Pr \left[ \abs{ X - \mu } > t \right] \le 2 \exp \left(- 2 t^2 / k \right)
\]
Setting $t = \sqrt{k \cdot \left( (n+2) \ln 2 + c \ln n\right)/2} $ gives
\[
\Pr \left[ \abs{ X - \mu } > t  \right] \le 2 \exp \left(- 2 t^2 / k \right)
\le \frac{1}{n^c 2^{n+1}}
\]

Let $\beta = |S| \cdot \sqrt{\left( (n+2) \ln 2 + c \ln n\right)/2}$.
We now apply this to two sequences of random variables, first for $H$ and then for $H^C$.
For each edge $e \in G[S]$, let $X_e$ denote the random variable for whether edge $e$ exists in the sampled graph $H$. Then for this sequence of $k:=E(S) \le |S|^2$ random variables, we get
\[
\Pr \left[ \abs{E_H(S) - p \cdot E(S)} > \beta \right]
\le \frac{1}{n^c 2^{n+1}}
\]
where we use the fact that $\beta$ is the same as $t$ with the setting of $k = |S|^2$. 
Now consider a new sequence of random variables, $X_e$ for each $e \not\in G[S]$, which denotes whether edge $e \in S \times S$ is chosen for the graph $H^C$, which happens with probability $1-p$ for each edge not in $G[S]$. For this sequence of $k := \binom{|S|}{2} - E(S)$ random variables, since $k \le |S|^2$ again, we get
\[
\Pr \left[ \abs{E_{H^C}(S) - (1-p) \cdot \left( \binom{|S|}{2} - E(S) \right)} > \beta \right]
\le \frac{1}{n^c 2^{n+1}}
\]
Recall that $\alpha = |S| \sqrt{(n + c \ln n)}/(2p-1)$, which we relate to $\beta$ as follows:
\[
\frac{\alpha \cdot (2p-1)}{\beta} 
= \sqrt{\frac{{n + c \ln n}}{{((n+2) \ln 2 + c \ln n) / 2}}}
= \sqrt{\frac{{2 n + 2c \ln n}}{{((n+2) \ln 2 + c \ln n)}}}
> 1
\]
where the last inequality holds for all $n \ge 2$.
Thus, since $\alpha > \beta/(2p-1)$,
\begin{align*}
\Pr \left[ \abs{\apx E(S) - E(S)} > 2\alpha \right]
&\le \Pr \left[ \abs{ E_{H}(S) - p \cdot {E(S)} } > (2p-1) \cdot \alpha \right] \\
&\ \ +
\Pr \left[ \abs{  E_{H^C}(S) - (1-p) \cdot \left( \binom{|S|}{2} - E(S) \right)  } > (2p-1) \cdot \alpha \right] \\
&\le \frac{1}{n^c 2^{n+1}} + \frac{1}{n^c 2^{n+1}}
\le \frac{1}{n^c 2^{n}}
\end{align*}
as required.
\end{proof}

\section{LEDP Low Out-Degree Vertex Ordering}

\cite{DLRSSY22} defines the \emph{low out-degree (vertex) ordering problem} to be finding a differentially private vertex ordering where
the out-degree of any vertex is minimized
when edges are oriented from vertices earlier in the ordering to later in the ordering. 
Specifically, there exists an ordering of the vertices of any input graph where
the out-degree is at most the degeneracy $d$ of the graph; the degeneracy of the graph 
is equal to its maximum $k$-core value. The previous best differentially private algorithm for 
the problem gives out-degree at most $(4+\eta)d + O\left(\frac{\log^3 n}{\eps}\right)$.
Here, we give a novel algorithm that obtains out-degree bounded by $d + O\left(\frac{\log n}{\eps}\right)$,
with high probability. 

\begin{algorithm}[h]
\caption{Low Out-Degree Vertex Ordering}
\label{alg:low-outdeg}
\textbf{Input:} Graph $G=(V,E)$, privacy parameter $\eps>0$.\\
\textbf{Output:} A vertex ordering $J$ with low out-degree in the induced edge orientation\\
\begin{algorithmic}[1]
\STATE $V_0\leftarrow V$, $t\leftarrow 0$, $k=c'\log{n}/\eps$.
\STATE Initialize $\hat{k}(v)\leftarrow 0$ for all $v\in V$
\STATE {\color{blue} Initialize ordered list $J$}
\FOR{$v\in V$}
\STATE $\tilde{\ell}(v)\leftarrow \text{Lap}(4/\eps)$
\ENDFOR{}
\STATE
\WHILE{$k\le n$}
\REPEAT
\STATE $t\leftarrow t+1$, $V_t\leftarrow V_{t-1}$
\FOR{$v\in V_{t-1}$}
\STATE $\nu_t(v)\leftarrow\text{Lap}(8/\eps)$
\IF{$d_{V_{t-1}}(v)+\nu_t(v)\le k+\tilde{\ell}(v)$}\label{low-outdegree-threshold}
\STATE $V_t\leftarrow V_t-\{v\}$
\STATE {\color{blue} Append $v$ to the end of $J$}
\ENDIF{}
\ENDFOR{}
\UNTIL{$V_{t-1}-V_{t} = \emptyset$}
\STATE
\STATE $k\leftarrow k + c'\log{n}/\eps$
\ENDWHILE{}
\STATE {\color{blue} \textbf{Release} $J$}
\end{algorithmic}
\end{algorithm}

Our algorithm (given in~\cref{alg:low-outdeg}) 
is a modification of~\cref{alg:optimal-algorithm} where each time we remove a 
vertex $v$ in Line~\ref{line:optimal peeled} of~\cref{alg:optimal-algorithm}, we add $v$ to the end of our current 
ordering. The changes to~\cref{alg:optimal-algorithm} are highlighted in blue.
Thus, our ordering is precisely the order by which vertices are removed. We now prove that 
our algorithm for low out-degree ordering is $\eps$-edge DP and gives an ordering for which the 
out-degree is upper bounded by $d + O\left(\frac{\log n}{\eps}\right)$.

\begin{theorem}\label{thm:low-out-deg}
    Our modified~\cref{alg:low-outdeg} is $\eps$-LEDP and gives with probability $1-1/poly(n)$ a low out-degree ordering
    where the out-degree is upper bounded by $d + O\left(\frac{\log n}{\eps}\right)$, where $d$ is the degeneracy of the graph, in $O(n)$ rounds.
\end{theorem}

\begin{proof}
    We first prove that our modified algorithm is $\eps$-LEDP. First, the proof of~\cref{thm:optimal-privacy} %
    ensures that the ratio of the probabilities that all vertices $v$ are removed at the same steps in edge-neighboring
    graphs $G$ and $G'$ is upper bounded by $e^{\eps}$. This is due to the privacy guarantee of MAT.
    For any sequence of MAT query outputs $\textbf{b}$ in Line~\ref{low-outdegree-threshold} of~\cref{alg:low-outdeg},
    the probability that the algorithm outputs $\textbf{b}$ for both $G$ and $G'$ differs by a factor of $e^{\eps}$.
    When vertices are added to $J$ is a post-processing of the outputs given by Line~\ref{low-outdegree-threshold}.
    Thus, since the ordering of the vertices follows
    precisely the order by which vertices are removed, then, the ordering that is released is $\eps$-LEDP.

    Now, we prove the upper bound on the out-degree returned by the algorithm. Each removed vertex $v$ has an induced degree that, when removed,
    is upper bounded by $k(v) + O\left(\frac{\log n}{\eps}\right)$, with high probability, by~\cref{thm:optimal-utility}. 
    The induced degree when the vertex is removed is exactly equal to its out-degree in the edge ordering induced by the vertex ordering $J$.
    Then, since $d = \max_{v \in V}\left(k(v)\right)$, the out-degree of any vertex in the edge orientation induced by $J$ is at most $d + O\left(\frac{\log n}{\eps}\right)$, with high probability.
\end{proof}

\subsection{LEDP Low Out-Degree Ordering in Low Rounds}\label{sec:low-out-ordering}

We now give a simple modification of our low round $k$-core decomposition algorithm to obtain a $\eps$-LEDP 
low out-degree ordering in $O(\log^2 n)$ rounds. As in the previous section, we highlight our minor changes in blue.
The only difference we make is to insert a tuple of the final level of each node and its index into an ordered list $J$.
Then, we return the sorted list where $J$ is sorted by level from smallest to largest level, breaking ties using vertex
ID.

\begin{algorithm}[htb!]
\caption{Improved $\eps$-LEDP Low Out-Degree Ordering in $O(\log^2 n)$ Rounds}
\label{alg:improved-outdegree}
\textbf{Input:} Graph $G=(V,E)$, privacy parameter $\eps>0$.\\
\textbf{Output:} A vertex ordering $J$ with low out-degree in the induced edge orientation\\
\begin{algorithmic}[1]
\STATE Set $\psi = 0.1\coren$ and $\lambda = \frac{2(30- \eta)\eta}{(\eta + 10)^2}$.\\
\FOR{$i = 1$ to $n$}
    \STATE $\tilde{\ell}(i) \leftarrow  \text{Lap}(4/\eps)$.
    \STATE $L_r[i] \leftarrow 0$ for all $r \in \left[\ceil{4\log^2 n}\right]$.
\ENDFOR
\FOR{$\lcur = 0$ to $\ceil{\numlevels} - 1$}
\FOR{$i = 1$ to $n$}
    \STATE $L_{\lcur + 1}[i] \leftarrow L_{\lcur}[i]$.\\
    \STATE $A_i \leftarrow 0$.
    \IF{$L_\lcur[i] = \lcur$}
        \STATE Let $\nup_{i}$ be the number of neighbors $j \in \adj_i$ ($\adj_i$ is $i$'s adjacency list) where $L_\lcur[j] =
                \lcur$.\\
        \STATE Sample $X \leftarrow  \text{Lap}(8/\eps)$.\\
        \STATE Compute $\hnup_{i} \leftarrow \nup_{i} + X$.\\
        \IF{$\hnup_{i} \leq \upexp^{\lfloor r/(\numgrouplevels)\rfloor} + \tilde{\ell}(i)$}\label{outdeg:abovethreshold}
            \STATE Stop all future queries for $i$ and 
            set $A_i \leftarrow 0$.
        \ELSE
            \STATE Continue and set $A_i \leftarrow 1$.\\
        \ENDIF
    \ENDIF
    \STATE $i$ \release $A_i$.\\
    \STATE $L_{\lcur + 1}[i] \leftarrow L_{\lcur}[i] + 1$ for every $i$ where $A_i = 1$.
\ENDFOR
\STATE Curator publishes $L_{\lcur + 1}$.\\
\ENDFOR
\STATE {\color{blue} $J \leftarrow \emptyset$}
\FOR{$i = 1$ to $n$}\label{outdeg:1}
    \STATE {\color{blue} Insert $J \leftarrow (L_{\ceil{\numlevels}-1}[i], i)$.}\label{outdeg:2}
\ENDFOR
\STATE {\color{blue} Sort $J$ first by level (first element in tuple) and then by index to break ties (second element in tuple)}\label{outdeg:3}
\STATE {\color{blue}\textbf{Release} $J$.}\label{outdeg:4}
\end{algorithmic}
\end{algorithm}

\begin{theorem}\label{thm:low-out-deg-low-round}
    Our low round~\cref{alg:improved-outdegree} is $\eps$-LEDP and gives, with probability $1-1/poly(n)$, a low out-degree ordering
    where the out-degree is upper bounded by $(2+\eta)\cdot d + O\left(\frac{\log n}{\eps}\right)$, where $d$ is the degeneracy of the graph, in $O(\log^2 n)$ rounds,
    for any constant $\eta > 0$.
\end{theorem}

\begin{proof}
    We proved in the proof of~\cref{lem:2-approx-ledp} that the released $L_r[i]$ values are $\eps$-LEDP. Hence, Lines~\ref{outdeg:1} to~\ref{outdeg:4} of~\cref{alg:improved-outdegree}
    perform post-processing on the released $L_r[i]$; hence,~\cref{alg:improved-outdegree} is also $\eps$-LEDP.

    Now we prove the out-degree returned by the algorithm. Each removed vertex $v$ has induced degree, that, when removed, is upper bounded by $k(v) \cdot
    (2 + \eta) + O(\eps^{-1}\log n)$; this is due to the fact that by~\cref{thm:approx-logn-rounds}, the algorithm outputs the approximate $k$-core using
    the value of the (first failed) above threshold query in Line~\ref{outdeg:abovethreshold}. Since the maximum $k$-core number is $d$,
    we obtain a $(2 + \eta)d + O(\eps^{-1}\log n)$ low out-degree ordering.
\end{proof} %
\section{LEDP Defective Coloring}

The standard greedy vertex coloring algorithm iterates through the vertices
and colors the vertices one-by-one in some arbitrary ordering. 
When coloring vertex $v$, it looks at colors already used by the neighbors $N(v)$ of $v$ and picks a color for $v$ 
which is not already used by one of the neighbors. We call such an unoccupied color a \emph{free} color. 
Colors used by neighbors are defined as \emph{used} colors.
A standard analysis shows that this greedy coloring algorithm
uses at most $\Delta+1$ colors, where $\Delta$ is the maximum degree.

In the differential privacy setting, we cannot directly check if a color is used by one of the neighbors of $v$, since the edges are private. 
Instead, we keep at every vertex $v$ and every color $c$ a count $\textsc{count}_v(c)$ indicating how many neighbors of $v$ are colored by $c$ and use MAT to check if a color has been used by more than $\Theta(\eps^{-1}\log{n})$ of the neighbors of $v$. 
We then choose a color for $v$ which has not been used by too many of the neighbors of $v$, as indicated by the output of MAT. Due to the utility guarantees of MAT (using the density function of the Laplace distribution), 
we know that the number of neighbors of a vertex $v$ which uses the same color as $v$ (i.e., the defectiveness of 
the coloring) is $\Theta(\eps^{-1}\log{n})$. Furthermore, following the standard analysis of 
greedy vertex coloring, we have that at most $O(\eps\alpha/\log{n})$  colors are used 
by this process, essentially matching the guarantees for differentially private coloring 
given by previous work \cite{christiansen2024private}.

To improve upon this, we first obtain a differentially private low out-degree ordering using \Cref{alg:low-outdeg}. Using the reverse of that ordering, we run the private greedy coloring algorithm described above, choosing colors based on the output of MAT. Since in this ordering, the in-degree of each vertex is bounded by $\alpha+O(\eps^{-1}\log{n})$ (where $\alpha$ is the degeneracy of the graph)
and the defectiveness remains $\Theta(\eps^{-1}\log{n})$, the algorithm uses an improved $O(\eps\alpha/\log{n}+1)$ colors. When the arboricity $\alpha$ is much smaller than the max degree, this improves upon previous work. We formally state the algorithm and prove its properties below.

\begin{algorithm}[t]
\caption{$\eps$-LEDP Vertex Coloring}
\label{alg:ledp-coloring}
\textbf{Input:} Graph $G=(V,E)$, privacy parameter $\eps>0$.\\
\textbf{Output:} $O(1+\eps\alpha/\log{n})$-vertex coloring with $O(\log{n}/\eps)$-defectiveness.\\
\begin{algorithmic}[1]
\STATE Run \Cref{alg:low-outdeg} with privacy parameter $\eps/2$ to obtain an ordering of the vertices $L$
\FOR{each vertex $v\in V$ and color $c\in\{1,\ldots,n\}$}
    \STATE set current count $\textsc{count}_v(c)\leftarrow 0$
    \STATE set noisy threshold $t_{v}(c)\leftarrow 100\log{n}/\eps + \text{Lap}(4/\eps)$
\ENDFOR
\FOR{$v\in V$ in the reverse order of $L$}
    \STATE set color $c(v)$ to be the smallest $c\in\{1,\ldots,n\}$ for which no neighbor of $v$ has output \textsc{threshold exceeded}

    \FOR{each neighbor $u\in N(v)$}
        \STATE increment $\textsc{count}_u(c(v))$ by 1
    \ENDFOR

    \FOR{each vertex $v\in V$ and color $c\in\{1,\ldots,n\}$}
        \IF{$\textsc{count}_v(c)+\text{Lap}(8/\eps)\ge t_v(c)$}           \STATE Node $v$ outputs: color $c$ \textsc{threshold exceeded} for node $v$
            \STATE Node $v$ stops answering future queries about color $c$
        \ENDIF
    \ENDFOR
\ENDFOR
\end{algorithmic}
\end{algorithm}

\begin{theorem}\label{thm:ledp-coloring}
    \Cref{alg:ledp-coloring} is $\eps$-LEDP. Furthermore, \Cref{alg:ledp-coloring} outputs an $O(\eps\alpha/\log{n}+1)$-vertex coloring with $O(\eps^{-1}\log{n})$-defectiveness with probability at least $1-O(1/n^2)$.
\end{theorem}
\begin{proof}
    First, we show that the algorithm is $\eps$-edge differentially private. We have that Line 1 is $\eps/2$-edge differentially private by \Cref{thm:low-out-deg}. The only part of the remainder of the algorithm which uses the private data is checking whether $\textsc{count}_u(c(v))$ exceeds the threshold in Line 12. We see that this is an instance of MAT where $\textsc{count}_u(c)$ are the queries for each color $c$ and vertex $v$, with privacy parameter $\eps/2$. In this instance of MAT, we see that the $\ell_1$ sensitivity $\Delta_1$ is 1 since the addition or removal of an edge $e=(u,v)$ with head $u$ and tail $v$ in our ordering only causes $\textsc{count}_u(c(v))$ to change by 1 and keeps all other counts unchanged, i.e, the count only changes for a single node, namely  $u$, and a single color, namely $c(v)$ (the change happens in Line 9). Hence, \Cref{lem:basic-comp} implies that the algorithm is $\eps$-differentially private.

    To see that the algorithm is in fact $\eps$-LEDP, we need to specify how MAT is implemented using local randomizers in the algorithm. The argument is similar to \Cref{thm:mat-ledp}, but it is slightly more general as each node has multiple thresholds. Each node $v$ stores its own $\textsc{count}_v(c)$ for each color $c$. Whenever a new node is colored, their color is output to the transcript. If the node is a neighbor of $v$, we update the corresponding $\textsc{count}_v(c)$. We then implement Lines 12--14 (even if the node isn't a neighbor of $v$). This gives the local implementation of the algorithm. It is easy to see that the local randomizers are themselves $\eps$-edge differentially private since their output is a subset of the output of the entire algorithm, and we have just shown that the algorithm is $\eps$-edge differentially private. Hence, the algorithm is $\eps$-LEDP.

    Next, we move to the utility guarantees. Observe that with probability at least $1-1/n^5$, we have that a Laplace random variable satisfies $|\text{Lap}(\beta)|\le 5\beta$. There are $O(n^3)$ Laplace random variables used in the entirety of the algorithm, so each of them satisfies $|\text{Lap}(\beta)|\le 5\beta$ with probability at least $1-1/n^2$ by the union bound. We will condition on this event for the remainder of the analysis.

    Observe that in order for the if-statement in Line 12 to evaluate to true, we must have $\textsc{count}_v(c)$ to be at least $40\log{n}/\eps$. This implies that for any node $v$, if some color cannot be used by it anymore, there must be at least $40\log{n}/\eps$ neighbors of $v$ which came before it in the ordering which used color $c$. Observe that the in-degree of each vertex based on the ordering is upper bounded by $ \alpha+O(\log(n)/\eps)$ by \Cref{thm:low-out-deg} with probability $1-n^{-O(1)}$. Since each color is used at least $40\log{n}/\eps$ times before its threshold is exceeded, this implies that the color $c(v)$ is at most $[\alpha+O(\log(n)/\eps)]/[40\log(n)/\eps]=O(1+\eps\alpha/\log{n})$
    in Line 7, so only $O(1+\eps\alpha/\log{n})$ colors are used in the algorithm.
    
    Furthermore, we have that if $\textsc{count}_v(c)\ge 160\log{n}/\eps$, Line 12 always evaluates to true and \textsc{threshold exceeded} is output for vertex $v$ and color $c$. When \textsc{threshold exceeded} is output for a vertex $v$ and color $c$, we have the property that no neighbors of $v$ will ever use color $c$ by definition in Line 7. This implies that the defectiveness is upper bounded by $160\log{n}/\eps$, completing the proof.
\end{proof}

\subsection{LEDP Coloring in Low Rounds}
Our $\eps$-LEDP vertex coloring algorithm that uses $\poly(\log n)$ number of rounds follows from the our greedy coloring algorithm given in~\cref{alg:ledp-coloring}
except for the following change. We combine~\cref{alg:ledp-coloring} 
with a procedure for assigning colors to different levels of the partition of vertices 
produced by our small round low out-degree ordering algorithm. This procedure is inspired by the non-private 
bounded degeneracy coloring algorithm of~\cite{HNW20}. We show the 
pseudocode for our procedure in~\cref{alg:coloring-low-rounds} with changes from~\cref{alg:ledp-coloring} indicated in blue.

Our algorithm first obtains a partitioning of vertices across levels using the low out-degree ordering algorithm given
in~\cref{sec:low-out-ordering}. Then, we give each level a \emph{different} set of colors to color the vertices in that level. We call each set of colors a \emph{palette}.
Let $C_i$ be the palette given to level $i$, then $C_i \cap C_j = \emptyset$ for any $i \neq j$. By the guarantees of our low out-degree ordering
algorithm, the size of the palette used in each level is upper bounded by $O(\alpha + \log n/\eps)$ since each node will have at most $O(\alpha \eps^{-1} \log n)$
neighbors in the same level. This results in a total of $O(\alpha\log^2 n + \log^3 n/\eps)$ palette colors.
Then, as in the previous section, we greedily color the nodes using MAT in the reverse order of the ordering returned except 
each node is colored using the palette associated with its level. The utility guarantee of MAT ensures the defectiveness is bounded by $O(\eps^{-1}\log n)$.
Finally, this results in at most $O\left(\eps\alpha\log n + \log^2 n\right)$ colors that are used, with high probability.

\begin{algorithm}[t]
\caption{Low Rounds $\eps$-LEDP Vertex Coloring}
\label{alg:coloring-low-rounds}
\textbf{Input:} Graph $G=(V,E)$, privacy parameter $\eps>0$.\\
\textbf{Output:} {\color{blue} $O(\log^2(n) + \eps\alpha \log{n})$-vertex coloring} with $O(\eps^{-1}\log{n})$-defectiveness.\\
\begin{algorithmic}[1]
\STATE Run \Cref{alg:improved-outdegree} with privacy parameter $\eps/2$ to obtain an ordering of the vertices $L$ {\color{blue} and a partitioning 
of vertices into $O(\log^2 n)$ levels}
\FOR{each vertex $v\in V$ {\color{blue} in level $i$ and color $c\in C^i = \{n \cdot i + 1, \dots, n \cdot (i + 1)\}$}}
    \STATE set current count $\textsc{count}_v(c)\leftarrow 0$
    \STATE set noisy threshold $t_{v}(c)\leftarrow 100\log{n}/\eps + \text{Lap}(4/\eps)$
\ENDFOR
\FOR{$v\in V$ in the reverse order of $L$}
    \STATE set the color $c(v)$ as the smallest {\color{blue} $c\in C_i$} for which no neighbor of $v$ {\color{blue} in level $i$} has output \textsc{threshold exceeded}

    \FOR{each neighbor $u\in N(v)$}
        \STATE increment $\textsc{count}_u(c(v))$ by 1
    \ENDFOR

    \FOR{each vertex $v\in V$ {\color{blue} in level $i$ with color $c\in C_i$}}
        \IF{$\textsc{count}_v(c)+\text{Lap}(8/\eps)\ge t_v(c)$}
            \STATE Node $v$ outputs: color $c$ \textsc{threshold exceeded} for node $v$ and stops answering future queries
        \ENDIF
    \ENDFOR
\ENDFOR
\end{algorithmic}
\end{algorithm}

\begin{theorem}\label{thm:ledp-coloring-low-rounds}
\cref{alg:coloring-low-rounds} is $\eps$-LEDP and returns a $O(\log^2(n) + \eps\alpha \log{n})$-vertex coloring with $O(\eps^{-1}\log n)$-defectiveness,
with probability at least $1 - O(1/n^2)$, in $O(\log^2 n)$ rounds.
\end{theorem}

\begin{proof}
    We first show that~\cref{alg:coloring-low-rounds} is $\eps$-LEDP by implementing the algorithm using MAT via local randomizers.
    As in the proof of~\cref{thm:ledp-coloring}, each node stores their count and noisy threshold in private memory. 
    Then, the color that is selected is output to the transcript. Each node increments their own private count using the 
    information on the transcript. This local randomizer is $\eps$-differentially private since given two neighboring 
    adjacency lists, the sensitivity of the count of the colors of adjacent neighbors is $1$. Hence, by the privacy of SVT, the local randomizer is 
    $\eps$-differentially private. We implement MAT by creating a vector of queries where each dimension is a color $c$ associated with a vertex $v$
    and each dimension contains the threshold queries of the count of $c$ held by $v$. In edge-neighboring graphs with edge $e^{*} = \{u, v\}$ that differs,
    the count of at most one color held by the endpoints differs between the two graphs. Thus, the sensitivity of our MAT queries is $2$ and 
    our algorithm is $\eps$-LEDP by~\cref{thm:mat-ledp}.

    Now we show the utility bounds. By the guarantees given for the out-degree of each vertex in the low out-degree ordering, 
    the maximum out-degree is $O(\alpha + \eps^{-1}\log n)$, with high probability. Hence, each node will require 
    a palette of size at most $O(\alpha + \eps^{-1}\log n)$. Now, we perform a similar analysis to that of the proof
    of~\cref{thm:mat-ledp}. We condition on the maximum noise being smaller than $40\eps^{-1}$ with probability at least $1 - 1/n^2$.
    In order for any query to evaluate to true, the number of colors must be at least $40\eps^{-1}\log n$. Each color must
    then be used by at least $40\eps^{-1}\log n$ neighbors on the same level. By the pigeonhole principle, this implies that
    the maximum number of colors used at any level is $O((\alpha + \eps^{-1}\log n)/\floor{40\eps^{-1}\log n}) = O(\alpha\eps/\log(n) + 1)$. 
    Since there are $O(\log^2 n)$ levels, the maximum number of colors used is $O(\alpha \eps\log(n) + \log^2(n))$. Furthermore, 
    the threshold query evaluates to true when the count exceeds $160\eps^{-1}\log(n)$ leading to a $O(\eps^{-1}\log n)$ defectiveness.
\end{proof}

\section{Acknowledgements}
\label{sec:acknowledgements}

\textit{Monika Henzinger and A. R. Sricharan}:
This project has received funding from the European Research Council (ERC) under the European Union's Horizon 2020 research and innovation programme (MoDynStruct, No. 101019564)  \includegraphics[width=0.9cm]{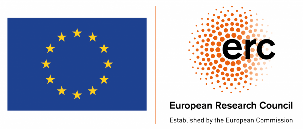} and the Austrian Science Fund (FWF) grant  \href{https://www.doi.org/10.55776/Z422}{DOI 10.55776/Z422} and grant  \href{https://www.doi.org/10.55776/I5982}{DOI 10.55776/I5982}. Laxman Dhulipala and George Z. Li are supported by NSF award number CNS-2317194. Quanquan C. Liu is supported by a Google Academic Research Award and by an NSF award number CCF-2453323.
 \bibliographystyle{alpha}
\bibliography{refs}

\appendix
\section{Related Work}\label{sec:relwork}
The  problems that we study
are closely related and have been studied side by side in the literature. For instance, the classic (non-private) $2$-approximation to the densest subgraph problem by Charikar~\cite{Charikar00} works by performing an iterative greedy peeling process; this procedure is similar to the 
classic greedy peeling procedure for core decomposition given by~\cite{MB83}. Subsequent work improved the approximation factor by using the multiplicative weight update framework~\cite{BGM14,SuVu20}. There has also been a large body of work on dynamic densest subgraph, with recent breakthrough results of~\cite{sawlani2020near,chekuri2024adaptive}
yielding $O(\mathsf{poly}(\log n))$ update time while maintaining a $(1+\eta)$-approximate densest subgraph.
The works of \cite{DLRSSY22,DKLV23} rely heavily on ideas from this literature.

On the privacy side, the first results for private densest subgraph are the works of Nguyen and Vullikanti~\cite{NV21} and
Farhadi, Hajiaghayi, and Shi~\cite{FHO21}. 
\cite{NV21} give algorithms which achieve a
$O(2 + \eta, \log(n) \log(1/\delta)/{\eps})$-approximation
under $(\eps, \delta)$-differential privacy. On the other hand, ~\cite{FHO21} give algorithms which achieve a
$(2+\eta,O({\log^{2.5} (n)\log(1/\sigma)}/{\eps}))$-approximation with probability $1-\sigma$
under $\eps$-differential privacy. More recently, \cite{DLRSSY22} gave a $(1+\eta, O(\mathsf{poly}(\log n)/\eps))$
approximation achieving $\eps$-differential privacy by adapting the muliplicative weight update distributed
algorithms of~\cite{BGM14,SuVu20} to the
private setting; however, the private algorithm is not distributed and runs in $O((n+m)\log^3 n)$ time.
Both~\cite{NV21} and~\cite{FHO21} provide
lower bounds on the additive error of any private densest subgraph algorithm; in particular, their lower bounds
state that any $\eps$-DP densest subgraph algorithm achieving high-probability accuracy guarantees
must have additive error $\Omega(\sqrt{\log(n)/\eps})$.
Our densest subgraph result yields the same multiplicative factor as~\cite{NV21, FHO21} but improves
the additive factor; on the other hand, we incur a worse multiplicative factor compared with~\cite{DLRSSY22}
but a better additive factor and our approximation is in the stronger
$\eps$-LEDP model.

For $k$-core decomposition, to the best of our knowledge \cite{DLRSSY22} is the first
work to study this problem; they give the first private algorithm achieving a $(2 + \eta, O(\mathsf{poly}(\log n)/\eps))$-approximation in the local edge-differential privacy setting by adapting a recently developed {\em parallel}
locally-adjustable dynamic algorithm from~\cite{LSYDS22}.
Since their approach is based on a parallel algorithm, improving on this result, e.g., to achieve any $(2-\eta)$-approximation seems
challenging due to a known $\mathsf{P}$-completeness result for degeneracy~\cite{anderson84pcomplete} (which reduces to $k$-core decomposition).
We take a different approach in this paper, and adapt a variant of the classical peeling algorithm~\cite{MB83}, which outputs exact core numbers with high probability. We note that this algorithm is {\em not parallelizable},
and indeed, the depth (i.e., longest chain of sequential dependencies) of this algorithm could be $\Omega(n)$ in the worst case. However, 
we give a low round algorithm for this problem at a increased $(2+\eta)$-approximation.

Finally, to the best of our knowledge, the only $\eps$-DP work on 
coloring is the very recent work of Christiansen, Rotenberg, Steiner,
and Vlieghe~\cite{christiansen2024private}. They introduce a 
$O(\Delta/\log(n) + \eps^{-1})$-coloring algorithm
with $O(\log n)$ defectiveness that is $\eps$-DP. We believe their 
algorithm could also be adapted to the $\eps$-LEDP model. Defectiveness
is defined as the number of colors a vertex shares with its neighbors.
Defectiveness in coloring 
is a well-studied property in the non-private vertex coloring 
community with many results in the central~\cite{D1987,BIMOR10,CowenGoddardJesurum97,
eaton1999defective,erdHos1967decomposition,
FRICK199345,Hendrey_Wood_2019,JING2022112637,
vskrekovski1999list,OssonaDeMendez2019,
HJWD2028,wood2018defective} and 
distributed~\cite{BarenboimElkin09,Kuhn09, BarenboimElkin11,FuchsKuhn23} settings.

\section{\texorpdfstring{Proof of the Approximation Factor in~\cref{sec:improved-distributed}}{Proof of the Approximation Factor in sec:improved-distributed}}\label{app:distributed-approx-proof}

\begin{proof}[Proof of~\cref{thm:approx-logn-rounds}]
    This proof is nearly identical to the proof of Theorem 4.7 in~\cite{DLRSSY22} except
    for two details. First, in~\cref{alg:improved-ledp-kcore}, we use Laplace noise instead
    of symmetric geometric noise and second, we add additional noise to the threshold when determining whether 
    vertices move up levels. Notice that these two changes \emph{only} affects Invariants 1 and 2. Thus, 
    we only need to show that modified versions of Invariants 1 and 2 still hold for our algorithm and 
    Theorem 4.7 and our approximation proof follow immediately.

    \begin{invariant}[Degree Upper Bound]\label{inv:degree-1}
    If node $i \in V_{\lcur}$ (where $V_\lcur$ is the set of nodes in level $\lcur$) and level $\lcur < \numlevels - 1$, then $i$
    has at most $\upexp^{\lfloor r/(\numgrouplevels)\rfloor} + \frac{c\log n}{\eps}$
    neighbors in levels $\geq r$ \whp{} for large enough constant $c > 0$.
    \end{invariant}
    
    \begin{invariant}[Degree Lower Bound]\label{inv:degree-2}
        If node $i \in V_{\lcur}$  (where $V_\lcur$ is the set of nodes in level $\lcur$) and level $\lcur > 0$, then
        $i$ has at least $(1 + \lf)^{\lfloor (r-1)/(\numgrouplevels)\rfloor} - \frac{c \log n}{\eps}$ neighbors in levels $\geq r - 1$ \whp{}
        for large enough constant $c > 0$.
    \end{invariant}

    First, we know that by the density function of the Laplace distribution, any
    noise drawn from the Laplace distributions used in our algorithm
    is upper bounded by $\frac{c\log{n}}{\eps}$ for large enough constant $c > 0$ \whp{}.
    We first show~\cref{inv:degree-1}. A node moves up a level when $\hnup_{i}$ exceeds the threshold given by 
    $\upexp^{\lfloor r/(\numgrouplevels)\rfloor} + \tilde{\ell}(i)$. By our algorithm, $\hnup_{i} = \nup_{i} + X$
    where $X$ and $\tilde{\ell}(i)$ are noises drawn from the Laplace distribution.
    Thus, if a node doesn't move up, then $\nup_{i} \leq \upexp^{\lfloor r/(\numgrouplevels)\rfloor} + \tilde{\ell}(i) - X$.
    The expression on the right is at most $\upexp^{\lfloor r/(\numgrouplevels)\rfloor} + 2 \cdot \frac{c\log{n}}{\eps}$
    \whp for large enough constant $c > 0$. 

    Now, we prove~\cref{inv:degree-2}. If a node is on level $> 0$, then it must have moved up at least one level.
    If it moved up at least one level (from level $r-1$), then its $\hnup_{i}$ at level $r-1$ must be 
    at least $\upexp^{\lfloor (r-1)/(\numgrouplevels)\rfloor} + \tilde{\ell}(i)$. Then, it holds that $\nup_i \geq 
    \upexp^{\lfloor (r-1)/(\numgrouplevels)\rfloor} + \tilde{\ell}(i) - X$ at level $r-1$. The right hand side
    is at least $\upexp^{\lfloor (r-1)/(\numgrouplevels)\rfloor} - 2 \cdot \frac{c\log n}{\eps}$ for large enough constant $c > 0$
    and~\cref{inv:degree-2} also holds.

    We now can use the proof of Theorem 4.7 exactly as written except with the new bounds given in~\cref{inv:degree-1}
    and~\cref{inv:degree-2} to obtain our final approximation guarantee.
\end{proof}

\end{document}